\documentclass[twocolumn,showpacs,superscriptaddress]{revtex4}
\usepackage{amsmath}
\usepackage{graphicx}

\newcommand{\nm}{\text{nm}}
\newcommand{\M}{\text{M}}
\newcommand{\mM}{\text{mM}}
\newcommand{\angstrom}{\text{\AA}}

\newcommand{\pH}{p\text{H}}
\newcommand{\pKa}{p\text{K}_\alpha}
\newcommand{\NPG}{N$_{\mathrm{PG}}$}
\newcommand{\CPG}{C$_{\mathrm{PG}}$}
\newcommand{\kB}{k_{\mathrm{B}}}
\newcommand{\kT}{\kB T}
\newcommand{\phiaz}{\phi^\alpha(z)}
\newcommand{\uaz}{u^\alpha(z)}
\newcommand{\ea}{\epsilon_\alpha}
\newcommand{\myav}[1]{\langle #1\rangle}
\newcommand{\q}[1]{q_{\mathrm{#1}}}
\newcommand{\sNC}{\sigma_{\mathrm{NC}}}
\newcommand{\sS}{\sigma_{\mathrm{s}}}
\newcommand{\pos}{{+}}
\renewcommand{\neg}{{-}}
\newcommand{\phiplus}{\phi_\pos}

\newcommand{\sfbox}{{\sc sfbox}}

\newcommand{\latin}[1]{{\itshape #1}}
\newcommand{\ie}{\latin{i.\,e.}}
\newcommand{\etal}{\latin{et al.}}

\newcommand{\lB}{l_{\mathrm{B}}}

\begin{document}

\title{Self-consistent field theory for the interactions
between keratin intermediate filaments}

\author{Anna Akinshina}

\affiliation{Unilever R\&D Port Sunlight, Quarry Road East, Bebington,
  Wirral, CH63 3JW, UK.}

\affiliation{School of Physics and Astronomy,
  University of Leeds, UK.}

\author{Etienne Jambon-Puillet}

\affiliation{Unilever R\&D Port Sunlight, Quarry Road East, Bebington,
  Wirral, CH63 3JW, UK.}

\affiliation{Institut Curie Centre de Recherche,
  CNRS UMR 168 - UPMC, F-75231, Paris Cedex 05, France.}

\author{Patrick B. Warren}

\affiliation{Unilever R\&D Port Sunlight, Quarry Road East, Bebington,
  Wirral, CH63 3JW, UK.}

\author{Massimo G. Noro}

\affiliation{Unilever R\&D Port Sunlight, Quarry Road East, Bebington,
  Wirral, CH63 3JW, UK.}

\date{December 3, 2012}

\begin{abstract}
We have developed a model for the interactions between keratin
intermediate filaments based on self-consistent field theory. The
intermediate filaments are represented by charged surfaces, and the
disordered terminal domains of the keratins are represented by charged
heteropolymers grafted to these surfaces. We estimate the system is
close to a charge compensation point where the heteropolymer grafting
density is matched to the surface charge density. Using a protein
model with amino acid resolution for the terminal domains, we find
that the terminal chains can mediate a weak attraction between the
keratin surfaces. The origin of the attraction is a combination of
bridging and electrostatics. The attraction disappears when the system
moves away from the charge compensation point, or when excess small
ions and/or NMF-representing free amino acids are added. These results
are in concordance with experimental observations, and support the
idea that the interaction between keratin filaments, and ultimately in
part the elastic properties of the keratin-containing tissue, is
controlled by a combination of the physico-chemical properties of the
disordered terminal domains and the composition of the medium in the
inter-filament region.
\end{abstract}



\maketitle

\section{Introduction}

The outermost layer of skin, the stratum corneum (SC), is often
described as organised into a `bricks-and-mortar' type structure,
where the mortar represents the self-assembled lipid lamellae and the
bricks refer to the protein-rich corneocytes.\cite{Mich75, Elias91,
  Sparr00, Lee09}.  Corneocytes are nonviable disk-shaped flat horny
cells mainly composed of keratin proteins, organised in complex
intermediate filament (IF) networks.  Keratins, in turn, are important
structural proteins which confer stiffness to many biological tissues
such as skin, nails and hair. There are 54 functional human keratin
genes, of which 28 are type I (acidic) keratin genes and 26 are type
II (neutral and basic) keratin genes.  A new systematic nomenclature
and functional role of keratins was presented by Schweizer
\etal\ \cite{Schweizer06}, Moll \etal\ \cite{Moll08}, and Gu and
Coulombe \cite{Gu07}.
  
\begin{figure}
\begin{center}
\includegraphics[width=2.85in]{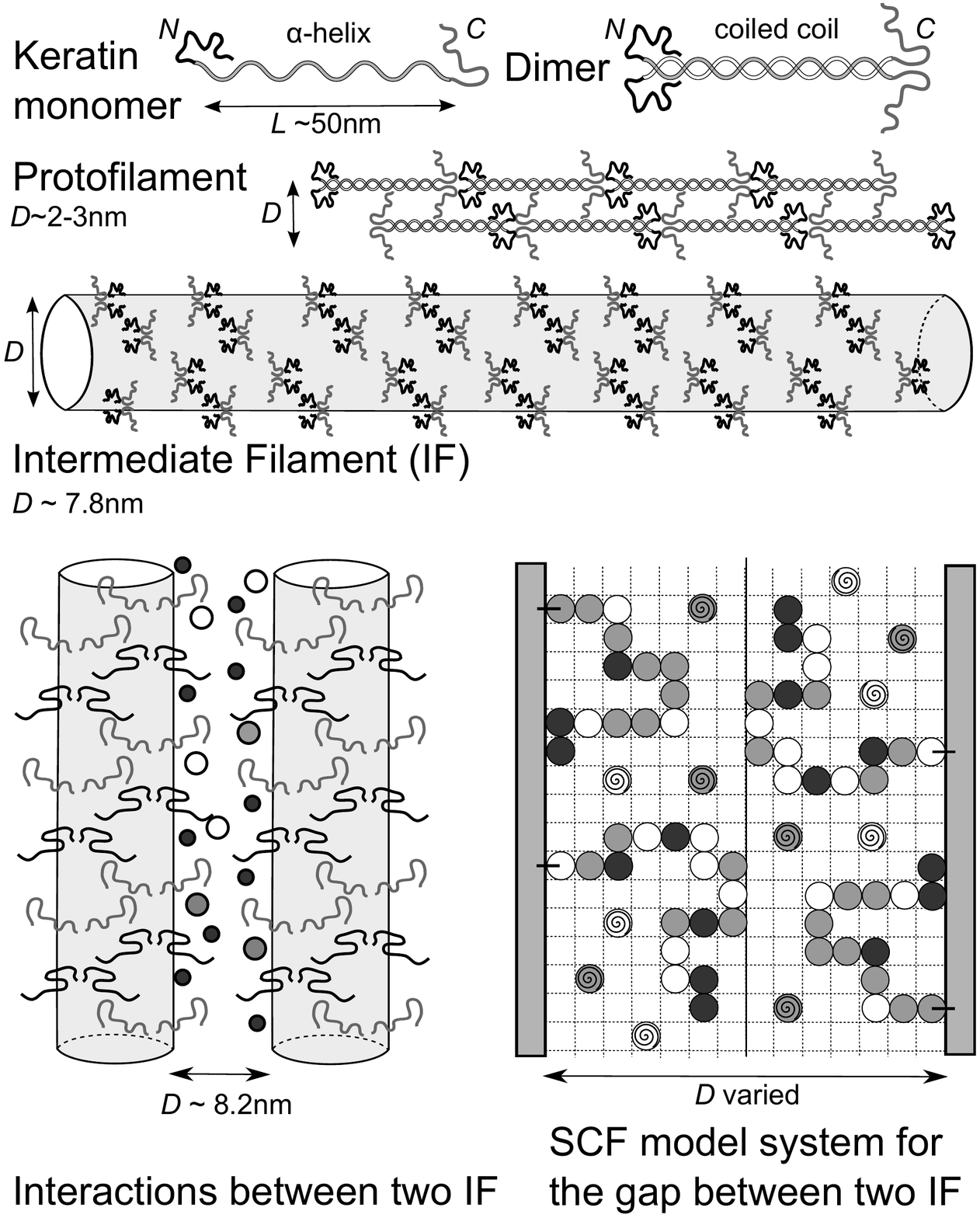}
\end{center}
\vskip -0.8in
\caption{IF organisation and the SCF model of N- and C-terminal
  domains attached onto IF surfaces. The IF surfaces are modelled as
  plane walls, the grafted domains as connected monomers, the salt
  ions and/or free amino acids as single monomers. All the other space
  is occupied by water. The separation between walls is varied in
  order to obtain the interaction potential mediated by the walls with
  grafted domains.\label{F1}}
\end{figure}

Keratin monomers consist of central $\alpha$-helical rod domains of
similar substructure ($\approx310$ amino acids) and two disordered
(unstructured) glycine-rich N- and C-terminal domains of variable
size.  Two keratin polypeptides associate in a parallel arrangement to
form an $\approx50\,\nm$ long coiled coil dimer, consisting of two
different type of keratins: one acidic (type I) and one neutral-basic
(type II). The most frequent keratin (K) dimer expressed in the SC and
the upper epidermis is the K1/K10 pair \cite{Stei91, Par94,
  Nor06}. The two coiled-coil heterodimers further self-assemble into
tetramers by packing into an antiparallel staggered
configuration. Tetramers, in turn, aggregate end-to-end forming
protofilaments with a diameter around 2--$3\,\nm$.  Two protofilaments
make a protofibril with diameter of order 4--$5\,\nm$; four of these
assemble laterally to form the keratin IF with diameter of order
8--$10\,\nm$ \cite{alberts-book, Fuchs90, Fuchs98, Nor04, Nor06,
  Nor07, Nor08, Schmuth07, Gu07,johnson-book}.  Schematically, a
keratin IF could be pictured as a long cylindrical object filled
mainly by $\alpha$-helical coiled coils domains, and decorated on the
surface by disordered N- and C-terminal domains extending into the
surrounding solution. An illustration of the IF hierarchical
organisation is depicted in Fig.~\ref{F1}.

Inside the corneocytes IFs are surrounded by a complex mixture of
water, ions, free amino acids and other low molecular weight water
soluble non-ionic compounds; this mixture is sometimes referred to as the
``Natural Moisturising Factor'' (NMF). NMF plays an important
role in skin moisturisation and in maintaining the physico-chemical
properties of the skin, such as elasticity and permeability
\cite{Jok95, Nak04, Hard86, Hard04, Rawl04, Tak12, Jac90, Rawl94}; and
it results from proteolytic degradation of filaggrin, a histidine-rich
protein \cite{Jok95, Nak04, Hard86, Hard04, Rawl04, Tak12, Jac90,
  Rawl94}.  A reduced amount of NMF correlates with dry, flaky and itchy
skin. Dry skin conditions may be a cosmetic problem triggered by
natural (seasonal) changes of SC physical properties \cite{Nak04,
  Hard04}, but they may escalate to severe inflammatory skin disorders
such as atopic dermatitis \cite{Hard04, Tak12}, xerosis \cite{Hard04,
  Jac90} ichthyosis \cite{Tak12, Rawl94} and psoriasis \cite{Tak12,
  Rawl94}.

Jokura \etal\ \cite{Jok95} studied the effect of NMF on SC elasticity
using NMR spectroscopy, rheology and electron microscopy. The authors
observed that treating an excised SC sample with water releases NMF
and leads to a keratin IF mobility reduction, and overall corneocyte
rigidity. Electron micrograph evidence suggested that in the absence
of NMF, keratin filaments tend to associated more tightly with each
other. Further hydration of the sample does not improve the mobility
of the fibers. However, the original IF mobility conditions were
partially restored by application of amino acid solutions. The authors
compared the effect of different types of amino acids on the
restoration of the SC elastic properties: neutral or basic amino
acids, such as glycine or lysine, provided remarkable recovery of SC
elasticity. In contrast, acidic amino acid, such as aspartic acid, was
not as effective.

These findings suggested the hypothesis that loss of SC elasticity is
due to increased intermolecular attractive forces between keratin
filaments. In physiological conditions, NMF plays the important role
to reduce these attractive forces, and to ensure SC elasticity.  It is
tempting to argue the protruding non-helical regions (unstructured N-
and C- domains) mediate the interaction between the NMF-rich matrix
and the IFs.  In this work we present a modelling study of the
interactions between keratin IFs suspended in different media: (i) a
salt free solution mimicking the NMF depleted system, and the effects of
(ii) added salt and (iii) NMF-rich amino acid solution.

\begin{figure}
\begin{center}
\includegraphics[height=7.25in]{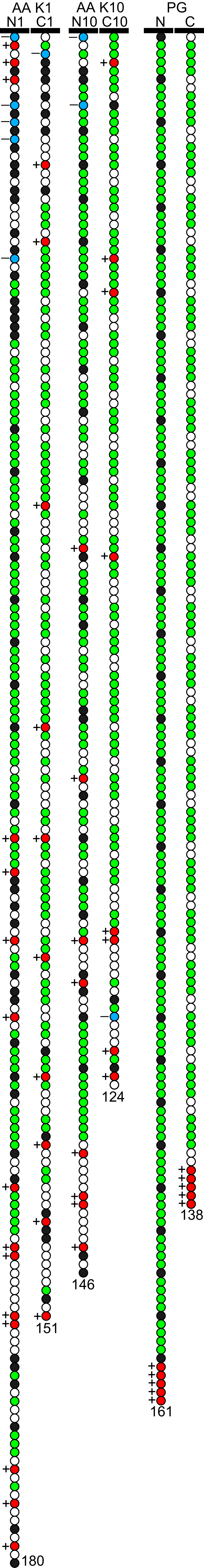}
\end{center}
\caption{Schematic illustration of the N and C unstructured terminal
  domains for the AA and PG models.\label{F2}}
\end{figure}

\section{Method}
\subsection{Self-consistent field (SCF) approach}
Interactions between two IFs formed by K1/K10 keratins were
investigated using the lattice self-consistent field (SCF) method
\cite{fleer-book, Israels93, Israels94, Bohmer90, Evers1, Evers2,
  Evers3}.  The helical cores of the two IF are modelled as planar
surfaces at distance $D$ apart with disordered N and C terminal
domains uniformly grafted onto them. The space between the IF surfaces
is filled by water molecules, ions and/or free amino acids. The
schematic model system is illustrated in Fig.~\ref{F1}.

In the lattice SCF scheme the space between two surfaces is divided
into layers $z = 1, 2, 3,\dots, D$ parallel to the walls, and each
layer is further divided into lattice cells of equal size. Each
lattice site is occupied by one of the monomeric species of the system
(i.e. by protein residue, water molecule, ion, etc.), so the total
volume fraction for all the species in each layer equals one,
$\sum_\alpha \phiaz=1$, where the volume fractions $\phiaz$ have the
meaning of dimensionless concentration of species type $\alpha$ at the
distance $z$ from the surface. Obtaining the equilibrium concentration
profiles for all the system components, $\phiaz$, is the primary
target of the SCF calculations. The volume fraction distributions
depend nonlinearly on the potential of mean force, $\uaz$, acting on
each species $\alpha$ in the system.  The potential for each
component, $\uaz$, in turn, depends on the volume fraction profiles,
as well as on the short range (Flory-Huggins) and long-range
(electrostatic) interactions between all the species of the system. To
find both quantities, $\phiaz$ and $\uaz$, a set of nonlinear
equations is constructed and solved self consistently by an iterative
procedure. The volume fraction profiles obtained in this way minimise
the free energy of the system \cite{Evers1}.

The SCF method is widely used to study properties of disordered
proteins at interfaces. Earlier, the scheme was implemented to
investigate adsorption of milk proteins, $\beta$-casein \cite{Leer96}
and $\alpha_{\mathrm{S1}}$-casein \cite{Dick97a, Dick97b, Ett08,
  Akin08}.  More recently the SCF approach was applied to protruding
terminal domains of neurofilaments (NF) \cite{Zhu07a, Zhu07b, Zhu09,
  Zhu10, Leer10a} and to microtubule-associated 3RS tau protein,
expressed in neurons of the central nervous system \cite{Leer10b}.
The detailed description of the method can be found in the original
literature. Here we apply the method to study unstructured terminal
domains of skin keratin IF.  We have considered and compared two
models for terminal N and C domains, detailed next.

\subsection{Amino acid (AA) model for terminal domains}
The first model (AA) is based on the primary structure of N and C
terminal domains for the keratins K1 and K10. The amino acid sequence
of these domains is taken from the Human IF Database \cite{if-note}. The
terminal domains of K1 (N1 and C1) consist of 180 and 151 amino acids,
respectively, and of K10 (N10 and C10) of 146 and 124 amino acids. All
the amino acids in this model are divided into five groups according
to their properties: `H', hydrophobic (Ala, Val, Leu, Ile, Met, Trp,
Phe, Pro, Cys); `P', polar (Ser, Thr, Tyr, Asn, Gln), `G' (Gly);
`$\pos$', basic (Arg, Lys, His); and `$\neg$', acidic (Glu, Asp). A
similar approach for allocating amino acids into groups is widely used
in literature \cite{Leer96, Dick97a, Dick97b, Ett08, Akin08, Zhu07a,
  Zhu07b, Zhu09, Zhu10, Leer10a, Leer10b}.

As can be seen from the amino acid sequence, N and C domains for K1
and K10 are glycine-rich ($\sim50$\%), where glycine is mostly
accumulated in blocks of 3--6 residues separated by one or two H or P
residues. Being aware that glycine is a peculiar amino acid, showing
both polar and hydrophobic behaviour (depending on the length of the
poly-glycine residue) \cite{Ohn06, Byk10, Noz71, Lu06}, we reserve for
glycine a separate classification group `G'.

The value of $\pH$ for SC is varied depending on SC depth, location
and environment and the reported values of SC $\pH$ are in the order
of 5--7 \cite{Hanson02, Parra03, Aberg08}.  According to $\pKa$
values, at $\pH = 7$ the amino acid residues Arg and Lys have charge
$q = +1\,e$, His has $q = +0.36\,e$, and Glu and Asp have $q =
-1\,e$. At $\pH = 5$, Arg and Lys have charge of $q = +1\,e$, His has
$q = +0.98\,e$, and Glu and Asp of $q = -0.76\,e$. In view of the
coarse grained level of the model, and to simplify the calculations,
we consider that each basic residue carries the charge of $q = +1\,e$
and each acidic residue of $q = -1\,e$. With such simplifications, the
total charge of N1 domain is $\q{N1} = +10\,e$ ($+15\,e$ and $-5\,e$),
for C1 it is $\q{C1} = +9\,e$ ($+10\,e$ and $-1\,e$), for N10 it is
$\q{N10} = +6\,e$ ($+8\,e$ and $-2\,e$), and the charge of C10 is
$\q{C10} = +7\,e$ ($+8\,e$ and $-1\,e$). The total charge of all four
terminal domains will be $\q{NC} = +32\,e$. The AA model for all four
domains is illustrated in Fig.~\ref{F2}.

\subsection{Polyglycine (PG) model for terminal domains}
When examining the central parts of the residue sequence in the AA
model in more detail, one can observe a repeating pattern of
polyglycine blocks separated by one or several H, P or, rarely, basic
monomers. One can also notice that the acidic residues are mostly
located at the beginning of the tails (near the helical IF part,
represented by planar surface in our SCF model) while the basic ones
are mostly situated at the end of the tails (far from the IF
surface). In order to capture and emphasize the major specific
properties of the terminal domain structure we have designed a
simplified ``polyglycine'' (PG) model for the N and C tails. The
coarse PG model consists of repeating blocks of four G monomers and
one H monomer (N tail) or four G and two P monomers (C tail) with
additional five basic residues at the end of each tail. Thus, the
structure of the N tail, \NPG,\ is H$_1$[G$_4$H$_1$]$_{31}(+)_5$ and
the structure of the C tail, \CPG,\ is
P$_1$[G$_4$P$_2$]$_{22}(+)_5$. The lengths of the \NPG\ and
\CPG\ fragments are chosen to be 161 and 138 residues, respectively:
this is because these numbers are near the average of N1 and N10 tail
lengths (180 and 146 residues) for \NPG\ and, consequently, the
average of C1 and C10 tail lengths (151 and 124 residues) for
\CPG. The \NPG\ and \CPG\ tail models are also illustrated in
Fig.~\ref{F2}.

\subsection{Modelling parameters}
The short-ranged Flory-Huggins interaction parameters $\chi$ between
the different types of monomers applied for both tail models are the
follows. The hydrophobic residues, H, strongly repel all the polar
ones, so we set $\chi = 2\,\kT$ for interactions of H with water and
ions (Na, Cl), and $\chi = 1\,\kT$ for interactions of H with all
the polar protein residues (P, $\pos$, $\neg$). The interactions of Na
and Cl with water are attractive, $\chi = -1\,\kT$ to mimic the
tendency of hydration for the ions. Concerning the last residue group,
G, we set $\chi = 0.4\,\kT$ for interactions of G and H group and
$\chi = 0.6\,\kT$ for those between G and all the others residues
(water, P, $\pos$, $\neg$, ions). The remaining interactions are set
to be athermal ($\chi = 0$). All the monomer types considered have no
affinity to the surface, $\chi_s=0$.

The choice of the interaction parameters for glycine is based on the
experimental data for solubility of free glycine and glycine
oligopeptides in water.  Experimental evidence show that free glycine
has rather good solubility in water \cite{Ohn06, Byk10, Noz71, Car99,
  Lu06}.  However, the solubility of oligoglycines is much lower and
it reduces with increase of the oligopeptide length \cite{Ohn06,
  Byk10, Lu06}.  Lu \etal\ \cite{Lu06} measured solubilities of
glycine and its oligopeptides up to hexaglycine at different $\pH$
values and the results show that the solubility of oligoglycines
longer than 3 residues strongly decreases with length. Bykov and Asher
\cite{Byk10} reported that oligoglycines longer than 5 residues are
normally insoluble in water; and Ohnishi \etal\ \cite{Ohn06} stated
that solubility of polypeptide with glycine linker beyond 6 is reduced
and polyglycine segments longer than 9 residues form insoluble
aggregates. In the current model, glycine is present in both forms: as
oligomers in the sequence of keratin terminal domains and as free
amino acid in the NMF composition. Taking into account dual
hydrophobic-hydrophilic properties of glycine, we anticipate that the
interactions of glycine with both hydrophobic and polar residues
should neither be strongly repulsive nor attractive. Thus, we set the
interactions with non-polar residues slightly attractive
($\chi=0.4\,\kT$) and with all polar slightly repulsive
($\chi=0.6\,\kT$).  Alternatively, it would be possible to separate
glycines into two groups: one for free glycine in NMF, and another one
for glycine blocks in terminal domains, but that is beyond the scope
of the current simplified model.

\newlength{\zeropointzero}
\settowidth{\zeropointzero}{0.0}
\newcommand{\zpz}[1]{\makebox[\zeropointzero][c]{#1}}

\begin{table}
\begin{center}
\begin{tabular}{c|ccccccccc|cc}
\hline
$\chi$ & \phantom{$-$}W & H & P & G & \zpz{$\pos$} & \zpz{$\neg$}
   & \zpz{Na} & \zpz{Cl} & \zpz{s} & $q$ &
$\epsilon_r$ \\
\hline
W      &  \phantom{$-$}-- &     &     &     &   &   &   &   &   & 0 & 80 \\
H      &  \phantom{$-$}2.0 &  --  &     &     &   &   &   &   &   & 0 & 2 \\
P      &  \phantom{$-$}0 & 1.0 &  --  &  &  & & &  &   & 0 & 5 \\
G      &  \phantom{$-$}0.6 & 0.4 & 0.6 &  --  &   &   &   &   &   & 0 & 4 \\
$\pos$ &  \phantom{$-$}0 & 1.0 &  --  & 0.6 & -- &   &   &   &   & $+1$ & 5 \\
$\neg$ &  \phantom{$-$}0 & 1.0 &  --  & 0.6 & -- & -- &   &   &   & $-1$ & 5 \\
Na     & -1.0 & 2.0 &  --  & 0.6 & -- & -- & -- &   &   & $+1$ & 5 \\
Cl     & -1.0 & 2.0 &  --  & 0.6 & -- & -- & -- & -- &   & $-1$ & 5 \\
s      &  \phantom{$-$}0 & --  &  --  &  --  & -- & -- & -- & -- & -- & (*) & 2 \\
\hline
\end{tabular}
\end{center}
\caption{Set of Flory-Huggins interaction parameters $\chi$ (in units
  of $\kT$), charges $q$ (in units of $e$), and relative dielectric
  permittivities $\epsilon_r$. A dash `--' indicates a zero entry. At
  (*) the surface (s) charge density (in units of $a_0^{-2}$) is
  varied between $\sS = -0.0655\,e$ and $\sS = -0.071\,e$ for the AA
  model (at $\sS = -0.0664\,e$ the surface charge is fully balanced by
  the charge of the grafted chains), and between $\sS = -0.0405\,e$
  and $\sS = -0.0425\,e$ for the PG model (at $\sS = -0.0415\,e$ the
  surface charge is again fully balanced by the charge of the grafted
  chains).\label{T1}}
\end{table}

We have considered different values of dielectric permittivities,
$\ea$, for different species components in our calculations.  A
similar approach has been used by Leermakers \etal\ for modelling
projection domains of neurofilaments \cite{Zhu07a, Zhu07b, Zhu09,
Zhu10, Leer10a}.  The permittivity for water was set to $\ea=80$,
for hydrophobic group H and IF surface $\ea = 2$, for all the polar
and charged components (P, $\pos$, $\neg$, Na, Cl) $\ea = 5$, and for
glycine (G) we set $\ea = 4$. The local dielectric permittivity was
calculated according to $\epsilon(z)=\epsilon_0\sum_\alpha\ea\phiaz$,
where $\epsilon_0$ is the permittivity of vacuum and $\phiaz$ is the
volume fraction of species type $\alpha$ at distance $z$.  The set of all
parameters for both AA and PG models is given in Table~\ref{T1}.

The calculations were carried out using the lattice spacing of $a_0 =
0.4\,\nm$. There are literature reports of lattice spacing ranging
between values of $0.3\,\nm$, used for modelling of caseins
\cite{Leer96, Dick97a, Dick97b, Ett08, Akin08} and $0.6\,\nm$ applied
for calculations of terminal domains of NF \cite{Zhu07a, Zhu07b,
  Zhu09, Zhu10, Leer10a} and 3RS tau protein \cite{Leer10b}. In our
model system the main components are amino acids and water
molecules. The water molecule size is about $3.1\,\angstrom$ and that
of amino acids ranges from $3.9\,\angstrom$ for Gly to
$6.1\,\angstrom$ for Trp \cite{trp-note}.  As glycine, the smallest
amino acid, is the main component of the sequence of the K1/K10
terminal domains, we used the intermediate value of $a_0 = 0.4\,\nm$
as lattice size in our calculations.

In this work we consider terminal domains uniformly grafted into
two IF cores, which are represented by planar surfaces. The grafting
density of the domains is calculated according to the fact that there
are four terminal domains (two N and two C) per one dimer length of $L
= 50\,\nm$, and one IF core consists of 16 dimers (8 protofilaments),
which gives in total 64 domains per dimer length. Taking the IF core
diameter of $2R = 7.8\,\nm$ \cite{Nor04} and the lattice size $a_0 =
0.4\,\nm$, we obtain the grafting density $\sigma = 64 a_0^2/(2\pi L R)
= 0.0083$ (in units of $a_0^2$).  For the AA model, with the average
charge per each tail $\myav{\q{N}} = \myav{\q{C}} = +8\,e$ (as the
total charge of the four domains is $\q{NC} = +32\,e$), the charge
density on the surface due to grafted chains would be $\sNC =
+0.0664\,e$. As for the PG model the charge density due to grafted chains
would be, correspondingly, $\sNC = +0.0415\,e$.

The calculation of the IF coiled-coil backbone charge is not so
obvious due to lack of information about IF core
organisation. Considering all the charged amino acids on the K1/K10
$\alpha$-helical parts, we have obtained $N_\pos = 45$, $N_\neg = 58$
for K1 and $N_\pos = 39$, $N_\neg = 60$ for K10, which gives the net
charge of the K1/K10 dimer $\q{dimer} = -34\,e$. We also take into
account that 14 salt bridges do not change the total charge of the
dimer. Carrying out the charge calculations for the IF core with dimer
length of $50\,\nm$, we consider, as previously, that IF backbone
comprises 16 dimers in its cross section. Thus, we obtained the
surface charge density (\ie\ charge per $a_0^2$) of $\sS = -0.071\,e$.
This result for the surface charge density is appeared to be quite
close to the value of the surface charge density due to the grafted
chains, $\sNC = +0.0664\,e$ (with opposite sign). As explained by
literature reports \cite{Fuchs90, Fuchs94, Fuchs98, Stein85a} the IFs
are apolar; we expect that the overall charge of IF core, dangling
terminal domains, and appropriate counterions should be
balanced. Because the exact IFs organisation is unknown, we can not
estimate how many of the accounted amino acids on the IF core are in
their dissociated form. Keeping in mind that some parts of the
protofilaments and, therefore, some of the charged amino acids could
be hidden inside the IF core where is no water, we presume that the
charge density of IF core could be lower then the calculated value of
$\sS = -0.071\,e$. Therefore, as a reference (starting) point for our
calculations for the AA model we consider the surface charge density
fully balanced with the charge density of the grafted chains, $\sS =
-0.0664\,e$. As for the PG model, the balanced value of the charge
density would be $\sS = -0.0415\,e$. We also explore a range of the
surface charge densities around these values for both AA and PG
terminal models.  Salt is represented by added Na and Cl ions and the
concentration was varied from as low as $c_s = 10^{-5}\,\M$ ($\phi_s =
3.8\times10^{-7}$) to $c_s = 0.1\,\M$ ($\phi_s = 3.8\times10^{-3}$),
depending on the system.

All the calculations were performed using the SCF code \sfbox\ kindly
provided by Frans Leermakers.

\begin{figure*}
\begin{center}
\includegraphics[width=5.25in]{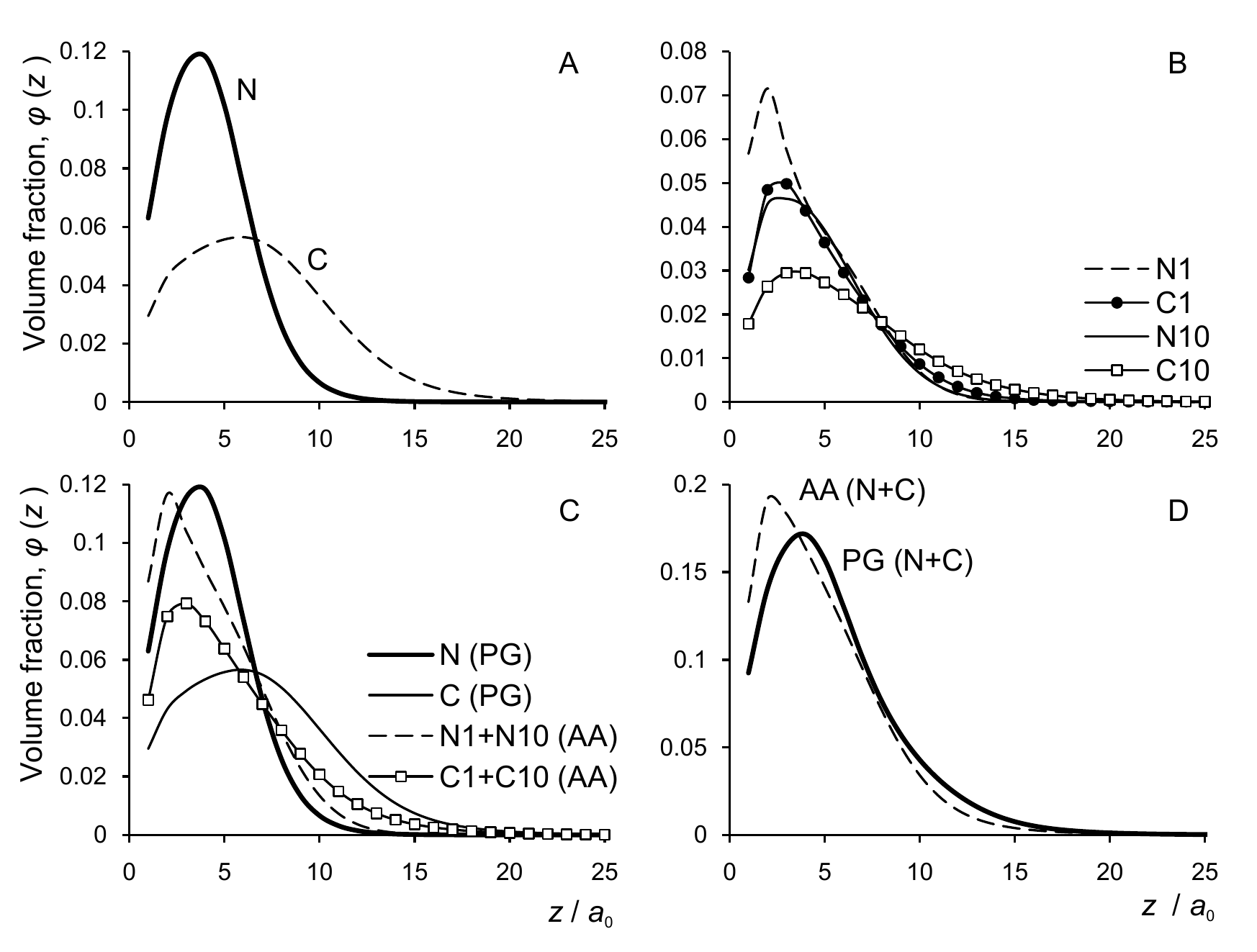}
\end{center}
%
%
\caption{Volume fraction profiles for (A) AA model, (B) PG model, (C)
  comparison of volume fraction for both N tails (N1+N10) and both C
  tail (C1 and C10) for the AA model with N and C tails of the PG
  model, and (D) volume fraction of all the tails for the AA and PG
  models.\label{F3}}
\end{figure*}

\section{Results and discussion}
\subsection{Overview}
Structural organization of the N and C terminal domains in one surface
and interactions between the two IF surfaces will be presented and
discussed below. The results considered include volume fraction
profiles, $\phi(z)$, of N and C terminals for both models,
corresponding profiles of the basic residues, $\phiplus(z)$, and
free energy of interactions between the two IF surfaces with the
attached terminal domains, $V(D)$. We start with the case of balanced
charge densities for the surface and chains, $|\sS|=\sNC$, at low ionic
strength, $c_s = 10^{-5}\,\M$. Then we consider the effect of added
salt and discuss the options when the surface charge in absolute value 
is higher or lower than the charge on the grafted chains. At the
end, in order to obtain better insights into the properties of each type of
terminal domains, we consider the interactions of the surfaces with only
one type of the chains (N or C) grafted.

\subsection{IF surfaces and tails at equal absolute charge: 
volume fraction profiles}
\begin{figure}
\begin{center}
\includegraphics[width=2.85in]{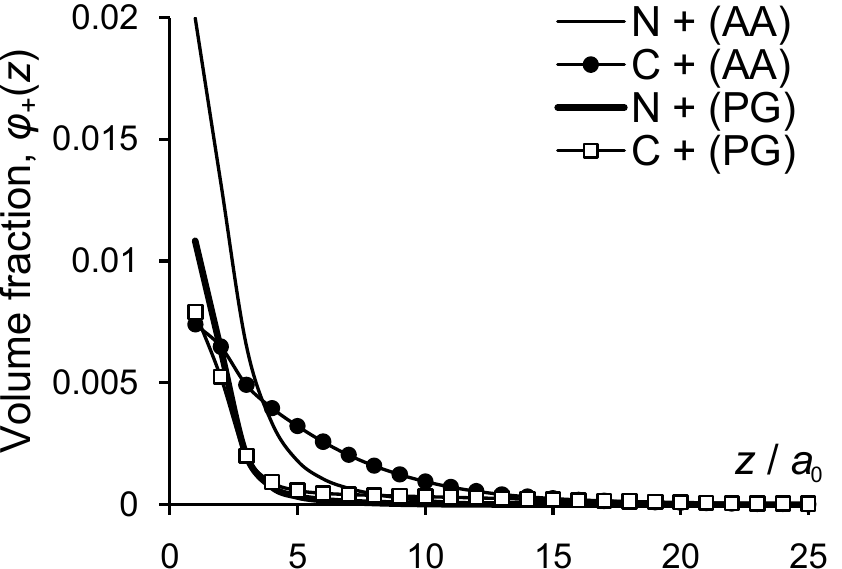}
\end{center}
\caption{Volume fraction of the basic residues on the N and C tails
  for the AA and PG models.\label{F4}}
\end{figure}

The volume fraction profiles, $\phi(z)$, show the monomer density of
the grafted N and C domains at distance $z$ from the
surface. These distributions provide an estimate of how far the grafted
chains extend from the surface and what is the most probable location
of any specified monomers. The volume fraction profiles for the whole
N and C chains are given in Fig.~\ref{F3} while Fig.~\ref{F4}
shows the distributions for only positively charged monomers of N and
C domains, $\phiplus(z)$. In order to obtain the spatial distribution
for an unaffected N and C chains, the profiles were obtained at large
surfaces separation, so that the grafted chains do not interact (this 
corresponds to the limit of an isolated IF in solution).

\subsubsection{Distribution of N and C tails}
In Fig.~\ref{F3}(A) we present the volume fraction profiles for the PG
model of N and C tails. The profiles for the two tails are quite
different: the monomer distribution of the more hydrophobic N tails is
more narrow compared with the profile for C tails, with most of the
monomers located in the first 10 layers from the surface and the
maximum density at $z = 4\,a_0$. The extension of the N tails does not
exceed $z = 13\,a_0$. More polar C tails have lower density near the
surface and more extended profiles. The maximum density is slightly
shifted away from the surface, $z = 6\,a_0$, and the profiles extend
up to $z = 20\,a_0$. With the simple block-copolymer model for
terminal domains we obtained the two distinct populations of the
chains: (i) more hydrophobic N tails are collapsed near the surface
and (ii) more polar C tails are projected farther into the
solution. However, we should notice that both types of chains are
actually quite compact near the surface. With contour lengths of
$161\,a_0$ and $138\,a_0$ the chains do not spread out more than
$13\,a_0$ and $20\,a_0$ respectively. We attribute this behaviour not
only to the hydrophobic nature of both tails, major component of which
is glycine, but also to the attraction of the positively charged
end-monomers to the negatively charged surface, causing formation of
loops.

The profiles for more realistic (AA) model for terminal domains,
presented in Fig.~\ref{F3}(B), show much smaller difference between
the distributions for N and C tails. In general, the behaviour of the
all four chains is similar: the distributions are quite narrow; most
of the monomers are located within the first 10 layers from the
surface, with the maximum density at $z = 2$--$3\,a_0$. The heights of
the density maxima reflect chain lengths, with the highest maximum for
the longest N1 tail and lowest one for the shortest C10. Having the
contour length of 124--$180\,a_0$ all the chains are in collapsed
state and do not protrude far into the solution due to their
hydrophobicity ($\sim50$\% of glycine) and the electrostatic
attraction to the surface. The highest value of volume fraction is
obtained for N1, the longest domain ($180\,a_0$). With the maximum in
layer 2, the chains do not extend more than $z = 14\,a_0$. The high
monomer density near the surface reflects the strong hydrophobic
properties of N1 tail---with 24\% of non-polar and 40\% of glycine
residues the chains prefer to be in compact conformation, reducing
contacts with the polar solvent. Similar tail extension is observed
also for N10 except that the maximum density value is lower than that
for N1, because N10 chains are shorter and slightly less hydrophobic
(17\% of H monomers and 47\% of G). The volume fraction profiles for C
tails are slightly more extended than those for N tails. In
particular, the distribution for C10 tail extends farthest, up to $z =
20\,a_0$, and the monomer density near the surface is reduced. Even
though C tails also consist of about 50\% G residues, the fraction of
non-polar H monomers is much smaller, 9\% and 2\% for C1 and C10,
respectively. Being more polar than N tails, C tails extend a little
farther into the solution.

\begin{figure}
\begin{center}
\includegraphics[width=2.85in]{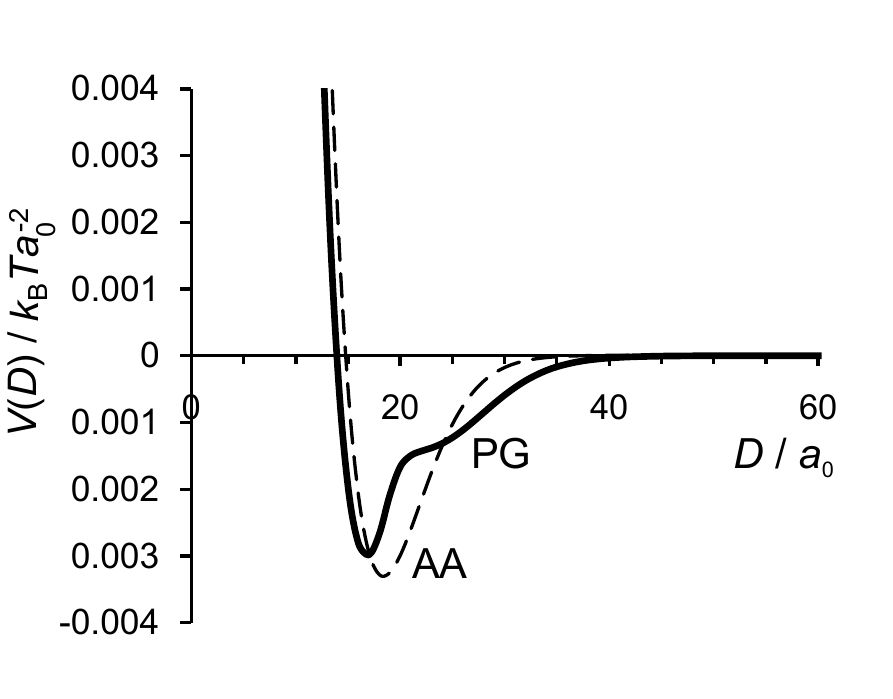}
\end{center}
\vskip -0.2in
\caption{Interaction potential $V(D)$ for the AA and PG models at the
  equal absolute charges of surfaces and grafted chains and low salt
  concentration, $c_s = 10^{-5}\,\M$.\label{F5}}
\end{figure}

In Fig.~\ref{F3}(C) and (D) we compare the monomer distributions for
the two models. Fig.~\ref{F5}(C) shows the profiles separately for N
and C tails and Fig.~\ref{F3}(D) compares the total profiles for N+C
tails together. The simplified PG model of the N tails gives the
density distribution quite similar to the combined profile for N1+N10
tails, see Fig.~\ref{F3}(C). Even though in the more detailed AA model
the maximum is slightly closer to the surface and the extension of the
profile is slightly larger (dashed line), these differences are
comparatively small. As for the C tails, the difference between the
profile for the PG model and the combined C1+C10 profile for the AA
model is more pronounced. The general shape of the profiles is
similar, so is their extension (to $z = 20\,a_0$), but the density
maximum for the PG model is lower and shifted away from the
surface. That gives the impression that the hydrophilicity of the C
tails in the PG model is somewhat overestimated; the more accurate AA
model predicts that the C tails are more hydrophobic. Nevertheless,
the relatively narrow profiles for the terminal domains coincide with
the prediction of the compact structure of the tails due to formation
of the glycine loops \cite{Stein91}.  The glycine loops hypothesis
predict that quasi-repetitive, glycine-rich terminal domains of
epithelial keratins comprise flexible and compact glycine loops, where
sequences of glycine make loops between the stacked non-polar
residues. Even though SCF method does not allow obtaining such
structural loops, it predicts compact conformation of the terminal
domains near the surface. Therefore, despite some discrepancies in
individual profiles for N and C tails, in the two models, the combined
profiles for all (N+C) tails are fully consistent with each other, see
Fig.~\ref{F3}(D). The simple glycine multi-block model for N and C
terminal domains reasonably well reflects the density distributions of
terminal domains for K1/K10 IF.

\subsubsection{Distribution of the basic residues of the tails} 
The volume fraction profiles for N and C domains provide the
information about spatial distribution of the chains as a whole, while
the location of the ends of the chains can be obtained from the
distribution of the positive residues, $\phiplus(z)$. The basic
residues in the PG model are located only at the end of the tails and
in the AA models they are scattered along the chains with higher
concentration near the ends. The volume fraction profiles for basic
residues in N and C tails, $\phiplus(z)$, are presented in
Fig.~\ref{F4}.

For the PG model the distributions of positively charged monomers for
N and C tails are practically the same. The positive monomers for both
tails are located near the surface, with maximum at the first layer
followed by abrupt decrease in the monomer density with the distance
from the surface. For the distances $z > 5\,a_0$ the fraction of basic
monomers becomes very small. That result allows us to conclude that
the basic residues, and, therefore, the end of the chains are located
at the surfaces, so the tails form either loops back to the grafting
surface or bridges with the opposite one.

As for the AA model, the distributions of positive residues for N and
C tails differ both from those in the PG model and between each
other. First, both distributions for the AA model are wider,
especially for the C tails, and second, the difference between the
$\phiplus(z)$ profiles for N and C tails is more noticeable. For the N
tails, $\phiplus(z)$ is similar to that for the PG model, with the
maximum at the first layer and subsequent decrease of the density with
distance. At distances $z > 10\,a_0$ very small fraction of basic
monomers can be found. The total volume fraction is higher than that
for the PG model because the amount of the positively charged monomers
is higher. In the PG model there are only 5 basic monomers in each
tail, while for N1 and N10 the numbers of basic monomers are 15 and 8,
respectively. Taking into account that the grafting density of N tails
for the PG model is the same as the sum of the grafting densities for
N1 and N10, the calculated total amount of the positive charges for
both N tails in the AA model is more than twice higher than that for
the PG model. That results in about double the value of volume
fraction of basic monomers for the AA model. Positively charged
monomers for C tails distribute much wider and spreading gradually
over $\sim17$ layers from the surface. The maximal density is again in
the first layer but its value is more than half than that for the N
tails, even though the number of positive charges for the C tails is
not much smaller, 9 and 8 for C1 and C10, respectively.

The density profiles for all the tails show that the maximum density
for the basic monomers is always at the first layer. The fact that the
highest concentration of those residues is at the surface confirm our
hypothesis that the charged monomers adsorb onto the surface, so the
chains form loops and/or bridges between the surfaces. Broader volume
fraction profiles of basic monomers for the AA model possibly results
from the different distribution of the charged monomers along the
chains. In the more detailed AA model, the basic monomers are not
located exactly at the end of the chains, but somehow distributed
along the whole length of the chains, with higher concentration at the
ends. Thereby, the more uniformly distributed charges in the AA model
give a thicker adsorbed layer while the clustered charges in the PG
model adsorb flat on the surface, producing a very thin layer, similar
to that of highly charged polyelectrolytes.

\begin{figure*}
\begin{center}
\includegraphics[width=5.25in]{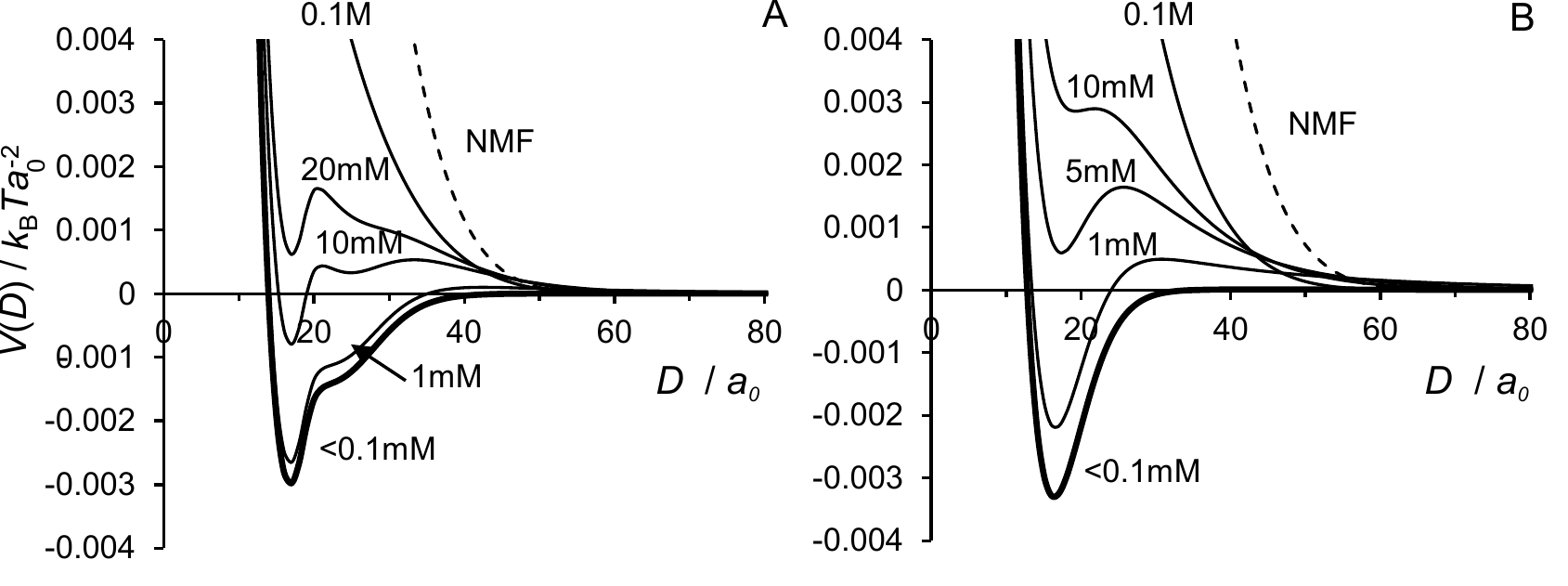}
\end{center}
\caption{Interaction potential $V(D)$ for (A) PG model, (B) AA model
  for various ionic strength indicated on the graphs at surface-chains
  charge balance conditions $|\sS|=\sNC$. Dashed lines represent data
  for NMF instead of salt.\label{F6}}
\end{figure*}

\subsection{IF surfaces and grafted tails at equal absolute charge: 
interaction potential profiles}
The interactions between the two surfaces (IF cores) covered by
grafted N and C terminal domains can be evaluated by calculating free
energy of interactions between the surfaces at each separation
$D$. The free energy of interactions $A(D)$ is calculated from the
partition function under conditions of restricted equilibrium,
described by Evers \etal\ \cite{Evers3}. Under such conditions some
components of the system are free to diffuse from the gap between the
two surfaces to the bulk solution (e.g. water molecules, ions, free
amino acids) and the others are restricted to stay within the gap
(e.g. grafted N and C domains). The net interaction potential, $V(D)$,
is the difference between the free energy value at separation $D$ and
its value when the surfaces are far apart, $V(D) = A(D)-A(D_\infty)$
and it is measured in units of $\kT/ a_0^2$. The ``far apart''
separation, $D_\infty$, is such that the two surfaces do not interact;
in our calculations $D_\infty$ ranges between $150\,a_0$ and
$1000\,a_0$, depending on the model and salt concentration. When the
interaction potential is negative, $V(D) < 0$, the two surfaces
attract each other, while the positive potential, $V(D) > 0$, implies
the repulsive interactions between the surfaces. It can be shown that
the interaction force between the two polymer-covered surfaces can be
evaluated from the obtained interaction potential \cite{Evers3,
  israel-book}.

\subsubsection{Low ionic strength}
The interaction potential for surfaces with attached N and C terminals
(for both models) in conditions of charge balance between surface and
chains, $|\sS|=\sNC$, is presented in Figs.~\ref{F5}
and~\ref{F6}. There is no need for additional counterions to satisfy
the charge neutrality condition.  Ideally we would run the SCF
calculation in the absence of added ions, but we are forced to
introduce an extremely small concentration of added salt to maintain
convergence. Still, the extremely low salt concentration case captures
the experimental set up of two IF surfaces immersed in deionised
water, where a small salt concentration cannot be avoided. Such is the
case of the Jukura experiment \cite{Jok95} where the water extractable
materials (NMF) from the SC sample were first released and then
deionised water was added. Other extreme cases, where the surface
charge is higher or lower than the charge on the terminal chains and
addition of certain amount of salt (counterions) is required to obtain
the charge neutrality, will be presented and discussed further below.

Fig.~\ref{F5} compares the interaction potential for both AA and PG
models at low salt content, $c_s= 10^{-5}\,\M$. For both models the
interaction potential has a well pronounced minimum at $D = 17\,a_0$
($6.8\,\nm$) for the PG model and $D = 18\,a_0$ ($7.2\,\nm$) for the
AA model, corresponding to net attractive interactions in the
system. It is interesting that the separations at which the attractive
minimum occurs ($D \approx 7\,\nm$) are in agreement with the
experimental values for the distance between the two IFs, $D \approx
8.2\,\nm$ \cite{Nor04}. The attraction between the surfaces at
separations $D \approx 15$--$35\,a_0$ (where $V(D) < 0$) occurs due to
the well known polyelectrolyte bridging effect \cite{Bohmer90, Evers3,
  SF85, Mik90} and favorable electrostatic conditions (ionic
strength). In the limit of low surface coverage, we believe that
positively charged end-monomers are attracted to the opposite surface
forming bridges across them. The possibility to be simultaneously
attracted to more than one surface is more entropically favorable. The
volume fraction profiles of the charged monomers discussed above
support this picture as the positively charged residues are mostly
located near the surface, which indicates the possibility of formation
either loops or bridges (if the surfaces are close enough). At larger
surface separations, $D > 35\,a_0$ the interaction potential
approaches zero, indicating that the grafted chains do not
interact. However, at short separations, $D < 15\,a_0$, the potential
is positive due to strong steric repulsion between the chains.

We should draw attention to the fact that our simple PG model for N
and C tails, based on the repetitive motif of glycine blocks, very
well reproduces the result of the more complex AA model based on the
amino acid sequence. Both characteristics of the system---the volume
fraction profiles and the interaction potential between the IF
surfaces---are in a good agreement between the two models. We believe
that the PG model can be slightly improved, for example, by
introducing some H residues into the C tail model and/or by
distributing the charge less blockwise along the chain. Despite its
simplicity, the PG model reflects well the properties of the N and C
domains and, therefore, it probably can be used as a starting point
for more refined (and computational intensive) modeling techniques,
such as MC, MD or DPD.

\subsubsection{High ionic strength and NMF}

The interaction potential for the two models at different ionic
strength is given in Fig.~\ref{F6}. As we already discussed, at low
salt content, $c_s < 0.1\,\mM$, the interaction potential develops an
attractive minimum at short separations between the surfaces and
levels to zero at longer separations. At higher ionic strength, $c_s
\approx 1$--$10\,\mM$, the minimum becomes shallower and a repulsion
apprears at larger separations. Our two models for terminal domains
give qualitatively similar results but in the PG model more added salt
is required to destroy the attraction, i.e. the PG model still shows a
small attraction at $c_s = 10\,\mM$, while for the AA model the
interactions are already repulsive at all separations at $c_s =
5\,\mM$.  That occurs because the attraction to the surface of the
charged block at the end of the PG chains is stronger than that of the
AA chains (more uniform charge distribution along the chains), so more
salt is needed to affect the attraction. In every case, at salt
concentration near physiological conditions, $c_s = 0.1\,\M$, the
strong repulsion between the surfaces with grafted chains is obtained
for both models.  High salt content leads to electrostatic screening,
so the grafted chain mediate essentially a steric repulsion between
the two opposing surfaces, much like polymer brushes.  Obviously, at
even higher ionic strength ($c_s>0.1\,\M$) the repulsion between the
surfaces becomes stronger and more short-ranged.

Calculations at varying levels of salt concentration were aimed at
mimicking the Jokura experiments with normal (healthy) and reduced
amounts of natural moisturizing factors (NMF) in SC.  The amount of
NMF is directly responsible for hydration level and elasticity of the
skin \cite{Hard86, Hard04, Rawl04, Jok95, Nak04}. NMF is made mostly
of free amino acids derived from the enzymatic degradation filaggrin,
as well as organic and inorganic salts \cite{Jok95, Nak04, Hard86,
  Hard04, Rawl04, Tak12, Jac90, Rawl94}.  As charged amino acids and
ions are important components of NMF, we have used high ionic strength
as the first approximation.  Increasing amounts of added salt tips the
IFs interaction from attractive to repulsive. Experiments show that
treatment with potassium lactate, could restore the SC hydration
\cite{Nak04}.

\begin{table}
\begin{center}
\begin{tabular}{l|r}
\hline
H & 13.7\,\% \\
P & 25.5\,\% \\
G & 10.5\,\% \\
$\pos$ & 8.4\,\% \\
$\neg$ & 11.2\,\% \\
$(+)^*$ & 2.8\,\% \\
water & 27.9\,\% \\
\hline
total & 100.0\,\% \\
\hline
\end{tabular}\\
{\rule{0pt}{12pt}\small $^*$Neutraliser.\qquad\qquad}
\end{center}
\caption{Composition of NMF solution.\label{T2}}
\end{table}

The next level of increasing complexity in our model was to account
for the complex mixture of amino acids in the suspending matrix
between the IFs. In order to create our coarse model of NMF we adopted
the amino acid composition form Jacobson \etal\ \cite{Jac90} and then
divided all amino acids into the same groups (H, P, G, $\pos$, $\neg$)
used in the model for N and C tails. The water content was set to 30\%
as reported in the literature \cite{Jok95, Rawl04, Warner88,
  Caspers01} and addition of neutraliser was necessary to ensure
charge neutrality in the bulk. The detailed composition of this NMF +
water model is given in Table~\ref{T2}.

The interaction potentials for two IF surfaces with grafted terminals
immersed into NMF solution are presented in Fig.~\ref{F6} by dashed
lines. The graphs clearly illustrate that the more complex NMF-water
mixture leads to an even stronger repulsion.  We have observed that
the mixture of free amino acids has stronger effect on the
interactions between IF surfaces than just adding salt to the solvent;
the NMF not only provides a strong repulsion between the approaching
surfaces but also ``pushes'' the surfaces further away from each
other. We believe the reason for such a strong repulsion between the
surfaces rests in the high amount of free charged species (ions and
amino acids), but what makes this forces more long ranged is the
presence of free neutral amino acids. Solution of only neutral amino
acids only slightly decreases the attraction between the surfaces and
shifts of the attraction minimum to larger separations.

This result does not support the Jokura \etal\ finding that neutral
amino acids improve mobility of keratin fibers, as well as basic amino
acids, but not acidic ones \cite{Jok95}. Our results showed that only
charged species in solution can affect the attractive intermolecular
forces between negatively charged IF cores with grafted positively
charged terminal chains. We should also mention, that in our
coarse-grained model we could not reveal the specific effect of basic
amino acids, as the properties of positive and negative free amino
acids are the same except of the charge and the charge neutrality is
required in the bulk. In order to examine the effect of specific ions
a more sophisticated model and/or method is required.

\begin{figure*}
\begin{center}
\includegraphics[width=5.25in]{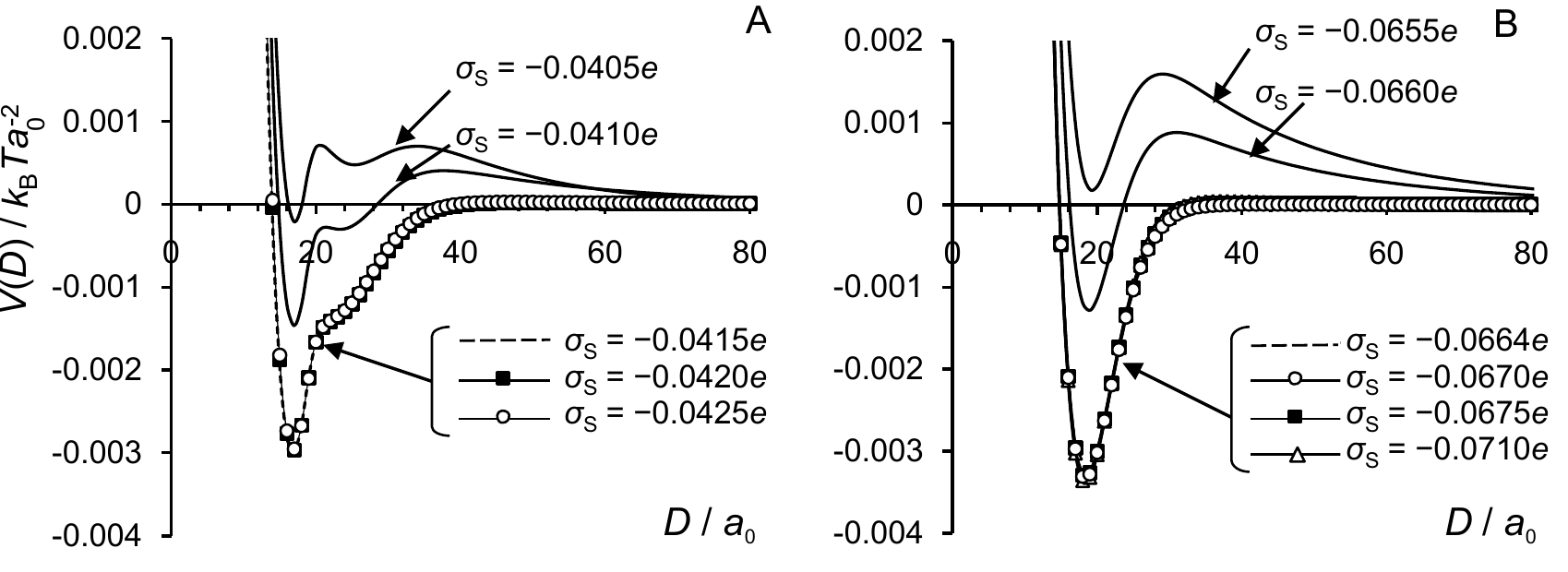}
\end{center}
\caption{Interaction potential for (A) PG model, (B) AA model for
  various surface charge indicated on the graphs neutralized by
  grafted tails and ions. The charge of the grafted chains is fixed
  via grafting density of $\sigma = 0.0083$. The amount of salt needed
  for charge neutrality depends on the surface charge and is given in
  Table~\ref{T3}.\label{F7}}
\end{figure*}

\subsection{Surface charge higher or lower than the charge of the
  terminals}
Previously we have described the case when charge on the surface is
fully balanced by charge on the grafted chains only. In this section
we consider cases when the charge of the surface is higher or lower
than the charge on the grafted tails, so to obtain charge neutrality
certain amount of counterions is needed.  We do so by setting the
concentration of salt $c_s$ in the bulk, which is in equilibrium with
the gap between the decorated IF surfaces.
 
In Fig.~\ref{F7} we present the interaction potential, $V(D)$, for the
surface charge densities of $\sS = -0.0655\,e$, $-0.0660\,e$,
$-0.0664\,e$, $-0.0670\,e$, $-0.0675\,e$, and $-0.0710\,e$ for AA
model and $\sS = -0.0405\,e$, $-0.0410\,e$, $-0.0415\,e$,
$-0.0420\,e$, and $-0.0425\,e$ for the PG model. The charge on the
grafted chains is kept constant at the values of $\sNC = 0.0664\,e$
for the AA model and $\sNC = 0.0415\,e$ for the PG model. Thus, for
the AA model, surfaces with charge density $|\sS| < 0.0664\,e$ are
``undercharged'' (in the specific sense that the surface charge
density is smaller in absolute value than that needed to balance the
charge on the grafted chains) and, correspondingly, with $|\sS| >
0.0664\,e$ they are ``overcharged''. For the PG model the threshold
values of surface charge density for undercharged and overcharged
surfaces would be, respectively, $|\sS| < 0.0415\,e$ and $|\sS| >
0.0415\,e$.

\newlength{\threeEthree}
\settowidth{\threeEthree}{$3{\times}10^{-3}$}
\newcommand{\tet}[1]{\makebox[\threeEthree][r]{#1}}

\begin{table}
\begin{center}
\begin{tabular}{ccc|ccc}
\hline
 & PG & & & AA & \\
\hline
$\sS(e)$ & $c_s(\M)$ & $\Delta\sigma$ & 
$\sS(e)$ & $c_s(\M)$ & $\Delta\sigma$ \\
$-$0.0405 & \tet{$3{\times}10^{-3}$} & \phantom{$-$}0.0010 &
   $-$0.0655 & \tet{$10^{-3}$} & \phantom{$-$}0.0010 \\
$-$0.0410 & \tet{$10^{-3}$} & \phantom{$-$}0.0005 &
   $-$0.0660 & \tet{$3{\times}10^{-4}$} & \phantom{$-$}0.0005 \\
$-$0.0415 & \tet{$10^{-5}$} & \phantom{$-$}0\phantom{.0000} &
   $-$0.0664 & \tet{$10^{-5}$} & \phantom{$-$}0\phantom{.0000} \\
$-$0.0420 & \tet{$5{\times}10^{-4}$} & $-$0.0005 &
   $-$0.0670 & \tet{$5{\times}10^{-4}$} & $-$0.0006 \\
$-$0.0425 & \tet{$10^{-3}$} & $-$0.0010 &
   $-$0.0675 & \tet{$10^{-3}$} & $-$0.0011 \\
 & & &
   $-$0.0710 & \tet{$4{\times}10^{-3}$} & $-$0.0046 \\
\hline
\end{tabular}
\end{center}
\caption{Values for surface charge density ($\sS$) and the
  corresponding salt concentration ($c_s$) required for charge
  neutralization.  The difference between the charge densities of the
  surface and grafted tails is denoted by $\Delta\sigma$.\label{T3}}
\end{table}

For each surface charge we found the values of the salt concentration
which provide charge balance. The surface charge densities with the
balancing salt concentrations are given in Table~\ref{T3}.

\begin{figure*}
\begin{center}
\includegraphics[width=5.25in]{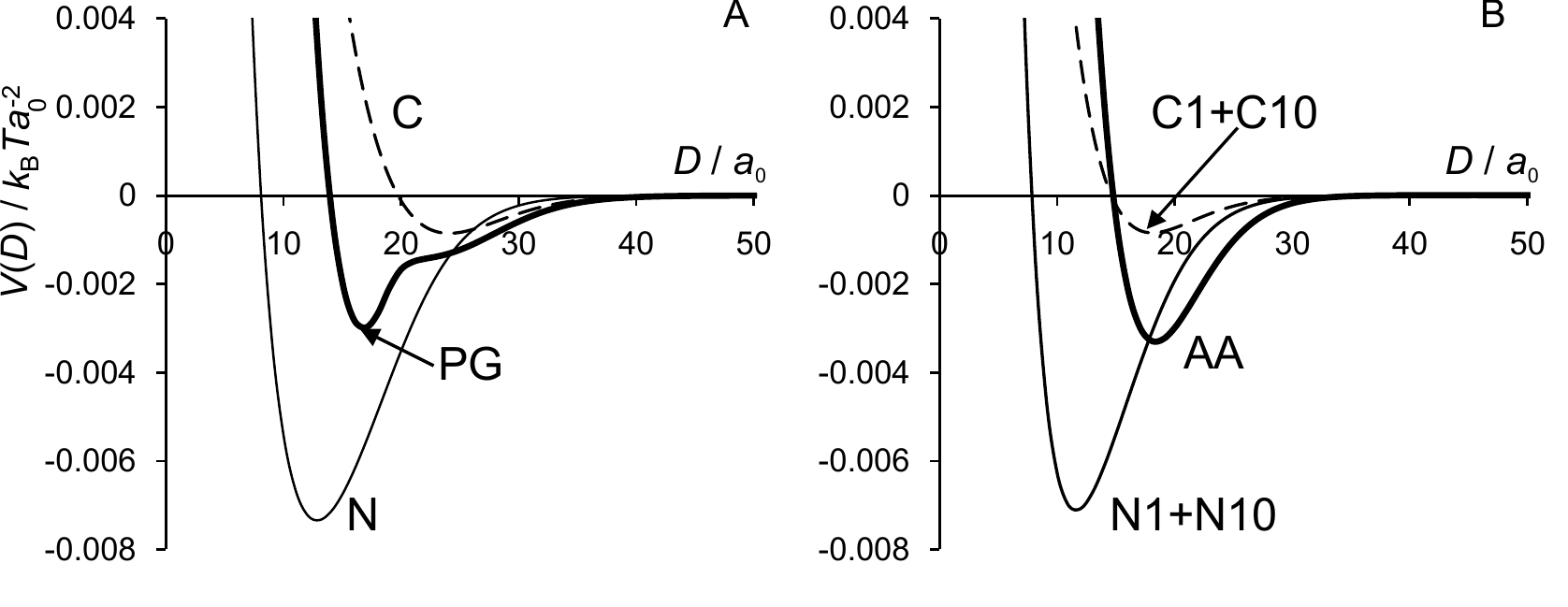}
\end{center}
\caption{Interaction potential for N and C terminal domains
  separately, (A) PG model; (B) AA model. The curves for the cases
  when all the tails are present are also given for comparison (bold
  lines).\label{F8}}
\end{figure*}

In the cases of under- or overcharged surface at low salt
concentrations, repulsive electrostatic forces dominate, so the
bridging attraction between the covered surfaces could not be seen. At
high ionic strength, the repulsion decreases due to screening. When
the surface is overcharged, i.e. when the charge on the surface is
higher in absolute value than the charge of the chains, it is possible
to find balancing salt concentration, under which the interaction
potential between the surfaces would be the same as for the case when
the surface charge is fully balanced by the charge of the chains only.
The graphs in Fig.~\ref{F7} show that at certain amount of added salt
the interaction potential profiles for $|\sS| \ge 0.0664\,e$ for the
AA model and for $|\sS|\ge 0.0415\,e$ for the PG model completely
overlap.  Table~\ref{T3} also shows that the stronger the charge
imbalance (difference between surface and chains charge), the higher
the amount of salt is required to neutralize the charge in the
system. However, when the surface is undercharged, the attractive part
is reduced and the potential always displays long-ranged repulsion,
which increases with increasing charge imbalance. This phenomenology
is the consequence of charge screening, as the following simplified
model calculation shows.  Consider a plane surface with a negative
surface charge density $\sS$, surmounted by a charge cloud at charge
density $\sNC$ uniformly distributed over a thickness $H$.  We solve
the linearised Poisson-Boltzmann equation for this problem,
\begin{equation}
\frac{d^2\varphi}{dz^2}-\kappa_s^2\varphi=\left\{
\begin{array}{cl}
-4\pi\lB\sNC/H & (0<z<H)\\
0 & (z>H)
\end{array}
\right.
\end{equation}
where $\varphi$ is the electrostatic potential (in units of $\kT/e$),
$\kappa_s$ is the inverse Debye screening length ($\kappa_s^2=8\pi\lB
c_s$), and $\lB$ is the Bjerrum length ($\lB\approx0.72\,\nm$).  The
boundary conditions are ${d\varphi}/{dz}=-4\pi\lB|\sS|$ at the wall
and $\varphi\to0$ as $z\to\infty$, and $\varphi$ should be continuous
at $z=H$ with a continuous first derivative.  This problem can be
solved analytically.  The behaviour of the potential at distances
$z>H$ from the surface is the relevant piece of information,
\begin{equation}
\varphi = \frac{\sNC \sinh(\kappa_s H)-|\sS|\kappa_s
  H}{2Hc_s}\times e^{-\kappa_s z} \quad(z>H)\,.
\end{equation}
The prefactor indicates there is a special balance point where the
potential vanishes completely for $z>H$.  This point occurs when
$|\sS|/\sNC=\sinh(\kappa_s H)/(\kappa_s H)$.  The right hand side is
an increasing function of $\kappa_s H$, and only approaches unity for
$\kappa_s H\to0$.  Thus we see the surface has to be overcharged in
order to reach the balance point and a higher degree of overcharging
requires a larger value of $\kappa_s H$ to compensate, corresponding
to higher salt, exactly as found above.  The reason for this is that
for $z\ge H$ the surface charge density is screened by an factor $\sim
e^{-\kappa_s H}$ relative to the diffuse oppositely-charged cloud.

The interactions between negatively charged surfaces covered by
positively charged polyelectrolytes were investigated experimentally
\cite{Clae92, Dahlgren93} by Monte-Carlo simulations
\cite{Dahlgren93}, and theoretically \cite{Borukhov99}. The results of
these studies have shown that the attractive bridging can dominate
only when the charges of the polymers and ions balance the charge of
the surface.  Claesson and Ninham \cite{Clae92} demonstrated that
attractive forces between mica surfaces covered by adsorbed chitosan
were observed only when electrostatic double layer disappeared,
i.e. when surface charges are exactly balanced by the charged of
adsorbed polysaccharide.  When charge of chitosan, controlled via
variation of solution $\pH$, was higher or lower than the charge of the
mica surfaces, the electrostatic double layer repulsion forces
dominate. Dahlgren \etal\ \cite{Dahlgren93} measured the force acting
between two mica surfaces covered by MAPTAC polyelectrolyte and also
carried out MC simulations for two surfaces with covered by oppositely
charged polyelectrolytes.  When PE adsorption was such that the
surface charge was balanced by the polyelectrolyte, a strong
attractive force was observed at short surface separations. Addition
of salt to the MAPTAC solution facilitates the increased adsorption of
polyelectrolyte, that leads to a reduced attraction and the appearance
of a repulsive double-layer force.  The authors concluded that the
attractive bridging mechanism will only dominate when the
polyelectrolyte adsorption approximately neutralizes the surface
charge density. Borukhov \etal\ \cite{Borukhov99} proposed a
theoretical approach to explain the behaviour of polyelectrolytes
between charged surfaces. Their calculations show that at low ionic
strength the attractive interactions between the surfaces take place
when polymer adsorption balances surface charge.  At high ionic
strength the surface charge is balanced both by polymers and ions and
the stronger the polymer charge, the more salt is needed to achieve
the charge neutrality. The authors also considered values of adsorbed
polymer higher or lower than the equilibrium adsorbed amount.  When
the adsorbed amount was lower than the equilibrium one, the attraction
was weaker. However, when the adsorbed amount was higher than the
equilibrium one, the results show stronger attraction between the
walls and also appearance of strong long-ranged repulsion, similar to
those shown in Fig.~\ref{F7}.

In the experiments described by Jokura \cite{Jok95} loss of elasticity
was observed for SC samples with extracted NMF and further hydrated by
addition of deionised water. The authors suggested that loss of
elasticity happens due to attractive intermolecular forces between
keratin fibers. NMF, mainly free amino acids, reduces intermolecular
forces through nonhelical regions of keratins (N and C terminal
domains), so the keratin filaments acquire their elasticity.  Our
modelling results, theoretical consideration and literature analysis
\cite{Clae92, Dahlgren93, Borukhov99} show that the attractive
interactions between IF at low salt content occur only when
$|\sS|=\sNC$ or the IF surfaces are slightly overcharged. We conclude
that the Jokura experiments could take place only at condition that
surface charge is equal or slightly higher than the charge on the
nonhelical chains. Thus, the charge of IF cores could not be much
higher or lower than the charge on the unstructured terminal domains.

\subsection{Role of each type of terminal domains}
Why has Nature used two types of unstructured terminal domains of
similar length scale for each keratin protein? Does each domain type
has a specific function and, if so, what is it?  Is it necessary to
capture the specific differences in a model?  Different authors have
taken different approaches. In modelling neurofilament projection
domains \cite{Zhu07a, Zhu07b, Zhu09, Zhu10, Leer10a}, the much shorter
globular N domains were not included in the study; only the C
projection domains were considered. On the contrary, in 3RS tau
protein research \cite{Leer10b} the authors focused on the 196 amino
acid long unstructured N domains. Thus, as the role of each domain in
keratins is yet unknown, in order to generate insights, we decided to
take advantage of fast computer models and examine the interactions
mediated by each type of terminal domains separately.

Fig.~\ref{F8} shows the interaction potentials for the IF cores
grafted only with N domains or only C domains. The graphs from
Fig.~\ref{F2} summarise the results and compare them against the full
model calculations. For the calculation of only one type of domains,
the grafting density of the chains was kept the same as before,
$\sigma = 0.00415$; this is half the total grafting density for both
chain types together ($\sigma = 0.0083$). The charge on the surface
was then adjusted to neutralize the charge from the chains, $\sS =
-0.0332\,e$ for the AA model and $\sS = -0.02075\,e$ for the PG model.

When only N chains are present, the minimum becomes much deeper and is
shifted closer to the surface. Even with the addition of $0.1\M$ of salt
this attraction minimum is still quite deep (data not
shown). Apparently, the more hydrophobic N tails behave as a ``glue'',
holding together the two surfaces.  In contrast, the C tails behave in
the opposite way. The interaction potentials for a similar model
including C-tails only result in much smaller attraction minimum,
pushed away from the surface. The more polar C-tails contribute much
less to the attraction between the IF surfaces.

It is tempting to propose that both N and C domains play important
roles in the structure and interactions of skin keratin IFs.  The more
hydrophobic N chains bring about a strong attraction between the IF
surfaces while the more polar C tails push the surfaces away from each
other, so that the two types of domains work together to keep IFs at
the optimal separation. Therefore, we believe that it is the
combination of both types of the domains balances the interactions
between the intermediate filaments.

\section{Conclusions}
We have applied the SCF approach to study interactions of the
unstructured N and C terminal domains of skin keratin (K1/K10)
Intermediate Filaments.  Positively charged N and C domains were
grafted onto negatively charged IF cores, represented by planar
surfaces. We have considered two models for N and C tails,
characterized by a different level of detail: the coarse
block-copolymer PG model and the more detailed AA model, which is
based on the amino acid sequence. In spite of the apparent simplicity
of the PG model, it qualitatively captures most of the effects
observed for the more complex AA model. We have presented monomer
density profiles for the N and C tails and, separately, profiles for
their basic residues only. We have compared and discussed interaction
potential profiles for IF surfaces with attached tails at various
surface charge densities, ionic strengths, and for the solution of
free amino acids representing NMF. We have also attempted to clarify
the role of each type of terminal domains considering N and C chains
separately. Our main findings are summarized as follows.
 
(A) Volume fraction profiles for N and C domains show that the
monomers of both types of the chains are mostly concentrated near the
surface, so the chain extension does not exceed $r\approx 20\,a_0 =
8\,\nm$ (Fig.~\ref{F3}). The basic residues of the terminal domains,
which are located near the end of the chains, have the highest density
at the (oppositely charged) surface (Fig.~\ref{F4}). These results
indicate that the chains form either loops or bridges with another
surface. Such bridges lead to attractive interactions between the two
IF surfaces at short separations. N tails are more hydrophobic and the
profiles for N tails are more narrow compared to those for C tails and
extend for no more than $r \approx 15\,a_0$ from the surface. The
interaction potential for surfaces covered by N domain type only
reveals that the attractive interactions between the surfaces are
stronger than those when both types of the domains considered together
and appear at shorter separations (Fig.~\ref{F8}). So we conclude that
N tails work as the ``glue'' between IF surfaces. C tails are slightly
more polar than N tails and extend slightly further into the solution
($r \approx 20\,a_0$). The interactions between surfaces with only C
tails grafted show much weaker and more long-ranged attraction. So, we
propose that C chains are ``responsible'' for keeping a certain
distance between IF. Hence, each type of the terminal domains has its
specific role and their combination retain IF at certain distance.

(B) When the charge of the IF surface is neutralized by the charge on
the grafted chains and the ionic strength is low, IFs experience
attractive force between each other at surface separations
$D\approx15$--$35\,a_0$ (6--$14\,\nm$) due to bridging effect of
grafted terminal domains. This attraction becomes weaker and turns
into repulsion with increase of ionic strength as a result of
electrostatic screening. The repulsion become stronger and longer
ranged when simple aqueous electrolyte solution between the IF
surfaces is replaced by a complex ``broth'' of amino acids---a coarse
grained representation of NMF in 30\% water. However, we can not
confirm experimental observations of Jokura \etal\ \cite{Jok95} that
neutral amino acids alone produce a similar effect. We have found that
charged small species such as ions or charged amino acids are
necessary components of NMF and their role is to decrease
electrostatic forces between IF. The effect of salt when ions differ
not only by their charge but also by size should be investigated using
more complicated model for salt molecules or/and by other simulation
methods.

(C) At low ionic strength the attraction between the IF surfaces can
be obtained only when the charge on the surface is fully compensated
by the charge on the chains and ions. That occurs only when the
surface charge is equal or slightly higher than the charge of the
grafted terminal domains. Therefore, we propose that: (i) negative
charge of the IF helical part is equal in absolute value or slightly
higher than the positive charge of the IF terminal domains; (ii) the
function of NMF is to prevent the attractive forces between protruding
terminal domains and IF helical cores. When NMF are removed or their
amount is highly reduced these attractive forces ``glue'' keratin
Intermediate Filaments and reduce the elasticity of the corneocytes.

The authors thank Chinmay Das, Peter Olmsted, Eugene Pashkovski, Jan
Marzinek, and Prof. Frans Leermakers for useful discussions.
Financial support from ERDF (grant B/680) and SoftComp (grant
RV(02)-11/390) is gratefully acknowledged.


\begin{thebibliography}{65}
\expandafter\ifx\csname natexlab\endcsname\relax\def\natexlab#1{#1}\fi
\expandafter\ifx\csname bibnamefont\endcsname\relax
  \def\bibnamefont#1{#1}\fi
\expandafter\ifx\csname bibfnamefont\endcsname\relax
  \def\bibfnamefont#1{#1}\fi
\expandafter\ifx\csname citenamefont\endcsname\relax
  \def\citenamefont#1{#1}\fi
\expandafter\ifx\csname url\endcsname\relax
  \def\url#1{\texttt{#1}}\fi
\expandafter\ifx\csname urlprefix\endcsname\relax\def\urlprefix{URL }\fi
\providecommand{\bibinfo}[2]{#2}
\providecommand{\eprint}[2][]{\url{#2}}

\bibitem[{\citenamefont{Michaels et~al.}(1975)\citenamefont{Michaels,
  Chandrasekaran, and Shaw}}]{Mich75}
\bibinfo{author}{\bibfnamefont{A.~S.} \bibnamefont{Michaels}},
  \bibinfo{author}{\bibfnamefont{S.~K.} \bibnamefont{Chandrasekaran}},
  \bibnamefont{and} \bibinfo{author}{\bibfnamefont{J.~E.} \bibnamefont{Shaw}},
  \bibinfo{journal}{AlChE Journal} \textbf{\bibinfo{volume}{21(5)}},
  \bibinfo{pages}{985} (\bibinfo{year}{1975}).

\bibitem[{\citenamefont{Elias}(1991)}]{Elias91}
\bibinfo{author}{\bibfnamefont{P.~M.} \bibnamefont{Elias}},
  \bibinfo{journal}{J. Controlled Release} \textbf{\bibinfo{volume}{15}},
  \bibinfo{pages}{199} (\bibinfo{year}{1991}).

\bibitem[{\citenamefont{Sparr and Wennerstr\"om}(2000)}]{Sparr00}
\bibinfo{author}{\bibfnamefont{E.}~\bibnamefont{Sparr}} \bibnamefont{and}
  \bibinfo{author}{\bibfnamefont{H.}~\bibnamefont{Wennerstr\"om}},
  \bibinfo{journal}{Colloids and Surfaces B: Biointerfaces}
  \textbf{\bibinfo{volume}{19}}, \bibinfo{pages}{103} (\bibinfo{year}{2000}).

\bibitem[{\citenamefont{Lee et~al.}(2009)\citenamefont{Lee, Ashcraft,
  Verploegen, Pashkovski, and Weitz}}]{Lee09}
\bibinfo{author}{\bibfnamefont{D.}~\bibnamefont{Lee}},
  \bibinfo{author}{\bibfnamefont{J.~N.} \bibnamefont{Ashcraft}},
  \bibinfo{author}{\bibfnamefont{E.}~\bibnamefont{Verploegen}},
  \bibinfo{author}{\bibfnamefont{E.}~\bibnamefont{Pashkovski}},
  \bibnamefont{and} \bibinfo{author}{\bibfnamefont{D.~A.} \bibnamefont{Weitz}},
  \bibinfo{journal}{Langmuir} \textbf{\bibinfo{volume}{25(10)}},
  \bibinfo{pages}{5762} (\bibinfo{year}{2009}).

\bibitem[{\citenamefont{Schweizer et~al.}(2006)\citenamefont{Schweizer, Bowden,
  Coulombe, Langbein, Lane, Magin, Maltais, Omary, Parry, Rogers
  et~al.}}]{Schweizer06}
\bibinfo{author}{\bibfnamefont{J.}~\bibnamefont{Schweizer}},
  \bibinfo{author}{\bibfnamefont{P.~E.} \bibnamefont{Bowden}},
  \bibinfo{author}{\bibfnamefont{P.~A.} \bibnamefont{Coulombe}},
  \bibinfo{author}{\bibfnamefont{L.}~\bibnamefont{Langbein}},
  \bibinfo{author}{\bibfnamefont{E.~B.} \bibnamefont{Lane}},
  \bibinfo{author}{\bibfnamefont{T.~M.} \bibnamefont{Magin}},
  \bibinfo{author}{\bibfnamefont{L.}~\bibnamefont{Maltais}},
  \bibinfo{author}{\bibfnamefont{M.~B.} \bibnamefont{Omary}},
  \bibinfo{author}{\bibfnamefont{D.~A.~D.} \bibnamefont{Parry}},
  \bibinfo{author}{\bibfnamefont{M.~A.} \bibnamefont{Rogers}},
  \bibnamefont{et~al.}, \bibinfo{journal}{J. Cell Biol.}
  \textbf{\bibinfo{volume}{174}}, \bibinfo{pages}{169} (\bibinfo{year}{2006}).

\bibitem[{\citenamefont{Moll et~al.}(2008)\citenamefont{Moll, Divo, and
  Langbein}}]{Moll08}
\bibinfo{author}{\bibfnamefont{R.}~\bibnamefont{Moll}},
  \bibinfo{author}{\bibfnamefont{M.}~\bibnamefont{Divo}}, \bibnamefont{and}
  \bibinfo{author}{\bibfnamefont{L.}~\bibnamefont{Langbein}},
  \bibinfo{journal}{Histochem Cell Biol} \textbf{\bibinfo{volume}{129}},
  \bibinfo{pages}{703} (\bibinfo{year}{2008}).

\bibitem[{\citenamefont{Gu and Coulombe}(2007)}]{Gu07}
\bibinfo{author}{\bibfnamefont{L.}~\bibnamefont{Gu}} \bibnamefont{and}
  \bibinfo{author}{\bibfnamefont{P.~A.} \bibnamefont{Coulombe}},
  \bibinfo{journal}{Curr. Opin. Cell Biol.} \textbf{\bibinfo{volume}{19}},
  \bibinfo{pages}{13} (\bibinfo{year}{2007}).

\bibitem[{\citenamefont{Steinert}(1991)}]{Stei91}
\bibinfo{author}{\bibfnamefont{P.~M.} \bibnamefont{Steinert}},
  \bibinfo{journal}{J Struct Biol} \textbf{\bibinfo{volume}{107}},
  \bibinfo{pages}{175} (\bibinfo{year}{1991}).

\bibitem[{\citenamefont{Paramio and Jorcano}(1994)}]{Par94}
\bibinfo{author}{\bibfnamefont{J.}~\bibnamefont{Paramio}} \bibnamefont{and}
  \bibinfo{author}{\bibfnamefont{J.}~\bibnamefont{Jorcano}},
  \bibinfo{journal}{Exp. Cell. Res.} \textbf{\bibinfo{volume}{215}},
  \bibinfo{pages}{319–} (\bibinfo{year}{1994}).

\bibitem[{\citenamefont{Norl\'en}(2006)}]{Nor06}
\bibinfo{author}{\bibfnamefont{L.}~\bibnamefont{Norl\'en}},
  \bibinfo{journal}{Int. J. Cosmetic Sci.} \textbf{\bibinfo{volume}{28}},
  \bibinfo{pages}{397} (\bibinfo{year}{2006}).

\bibitem[{\citenamefont{Alberts et~al.}(2002)\citenamefont{Alberts, Johnson,
  Lewis, Raff, Roberts, and Walter}}]{alberts-book}
\bibinfo{author}{\bibfnamefont{B.}~\bibnamefont{Alberts}},
  \bibinfo{author}{\bibfnamefont{A.}~\bibnamefont{Johnson}},
  \bibinfo{author}{\bibfnamefont{J.}~\bibnamefont{Lewis}},
  \bibinfo{author}{\bibfnamefont{M.}~\bibnamefont{Raff}},
  \bibinfo{author}{\bibfnamefont{K.}~\bibnamefont{Roberts}}, \bibnamefont{and}
  \bibinfo{author}{\bibfnamefont{P.}~\bibnamefont{Walter}},
  \emph{\bibinfo{title}{Molecular biology of the cell}}
  (\bibinfo{publisher}{Garland Science}, \bibinfo{address}{New York},
  \bibinfo{year}{2002}).

\bibitem[{\citenamefont{Coulombe and Fuchs}(1990)}]{Fuchs90}
\bibinfo{author}{\bibfnamefont{P.~A.} \bibnamefont{Coulombe}} \bibnamefont{and}
  \bibinfo{author}{\bibfnamefont{E.}~\bibnamefont{Fuchs}}, \bibinfo{journal}{J.
  Cell Biol.} \textbf{\bibinfo{volume}{111}}, \bibinfo{pages}{153}
  (\bibinfo{year}{1990}).

\bibitem[{\citenamefont{Fuchs and Cleveland}(1998)}]{Fuchs98}
\bibinfo{author}{\bibfnamefont{E.}~\bibnamefont{Fuchs}} \bibnamefont{and}
  \bibinfo{author}{\bibfnamefont{D.~W.} \bibnamefont{Cleveland}},
  \bibinfo{journal}{Science} \textbf{\bibinfo{volume}{279}},
  \bibinfo{pages}{514} (\bibinfo{year}{1998}).

\bibitem[{\citenamefont{Norl\'en and {Al-Amoudi}}(2004)}]{Nor04}
\bibinfo{author}{\bibfnamefont{L.}~\bibnamefont{Norl\'en}} \bibnamefont{and}
  \bibinfo{author}{\bibfnamefont{A.}~\bibnamefont{{Al-Amoudi}}},
  \bibinfo{journal}{J. Invest. Dermatol.} \textbf{\bibinfo{volume}{123}},
  \bibinfo{pages}{715} (\bibinfo{year}{2004}).

\bibitem[{\citenamefont{Norl\'en et~al.}(2007)\citenamefont{Norl\'en, Masich,
  Goldie, and Hoenger}}]{Nor07}
\bibinfo{author}{\bibfnamefont{L.}~\bibnamefont{Norl\'en}},
  \bibinfo{author}{\bibfnamefont{S.}~\bibnamefont{Masich}},
  \bibinfo{author}{\bibfnamefont{K.~N.} \bibnamefont{Goldie}},
  \bibnamefont{and} \bibinfo{author}{\bibfnamefont{A.}~\bibnamefont{Hoenger}},
  \bibinfo{journal}{Exp. Cell Res.} \textbf{\bibinfo{volume}{313}},
  \bibinfo{pages}{2217} (\bibinfo{year}{2007}).

\bibitem[{\citenamefont{Norl\'en}(2008)}]{Nor08}
\bibinfo{author}{\bibfnamefont{L.}~\bibnamefont{Norl\'en}},
  \bibinfo{journal}{Eur. J. Dermatol.} \textbf{\bibinfo{volume}{18}},
  \bibinfo{pages}{279} (\bibinfo{year}{2008}).

\bibitem[{\citenamefont{Schmuth et~al.}(2007)\citenamefont{Schmuth, Gruber,
  Elias, and Williams}}]{Schmuth07}
\bibinfo{author}{\bibfnamefont{M.}~\bibnamefont{Schmuth}},
  \bibinfo{author}{\bibfnamefont{R.}~\bibnamefont{Gruber}},
  \bibinfo{author}{\bibfnamefont{P.~M.} \bibnamefont{Elias}}, \bibnamefont{and}
  \bibinfo{author}{\bibfnamefont{M.~L.} \bibnamefont{Williams}},
  \bibinfo{journal}{Adv. Dermatol.} \textbf{\bibinfo{volume}{23}},
  \bibinfo{pages}{231} (\bibinfo{year}{2007}).

\bibitem[{\citenamefont{Johnson and Harley}(2011)}]{johnson-book}
\bibinfo{author}{\bibfnamefont{A.~W.} \bibnamefont{Johnson}} \bibnamefont{and}
  \bibinfo{author}{\bibfnamefont{B.~A.~C.} \bibnamefont{Harley}},
  \emph{\bibinfo{title}{Mechanobiology of cell-cell and cell-matrix
  interactions}} (\bibinfo{publisher}{Springer}, \bibinfo{address}{New York},
  \bibinfo{year}{2011}).

\bibitem[{\citenamefont{Jokura et~al.}(1995)\citenamefont{Jokura, Ishikawa,
  Tokuda, and Imokawa}}]{Jok95}
\bibinfo{author}{\bibfnamefont{Y.}~\bibnamefont{Jokura}},
  \bibinfo{author}{\bibfnamefont{S.}~\bibnamefont{Ishikawa}},
  \bibinfo{author}{\bibfnamefont{H.}~\bibnamefont{Tokuda}}, \bibnamefont{and}
  \bibinfo{author}{\bibfnamefont{G.}~\bibnamefont{Imokawa}},
  \bibinfo{journal}{J. Invest. Dermatol.} \textbf{\bibinfo{volume}{104}},
  \bibinfo{pages}{806} (\bibinfo{year}{1995}).

\bibitem[{\citenamefont{Nakagawa et~al.}(2004)\citenamefont{Nakagawa, Sakai,
  Matsumoto, Yamada, Nagano, Yuki, Sumida, and Uchiwa}}]{Nak04}
\bibinfo{author}{\bibfnamefont{N.}~\bibnamefont{Nakagawa}},
  \bibinfo{author}{\bibfnamefont{S.}~\bibnamefont{Sakai}},
  \bibinfo{author}{\bibfnamefont{M.}~\bibnamefont{Matsumoto}},
  \bibinfo{author}{\bibfnamefont{K.}~\bibnamefont{Yamada}},
  \bibinfo{author}{\bibfnamefont{M.}~\bibnamefont{Nagano}},
  \bibinfo{author}{\bibfnamefont{T.}~\bibnamefont{Yuki}},
  \bibinfo{author}{\bibfnamefont{Y.}~\bibnamefont{Sumida}}, \bibnamefont{and}
  \bibinfo{author}{\bibfnamefont{H.}~\bibnamefont{Uchiwa}},
  \bibinfo{journal}{J. Invest. Dermatol.} \textbf{\bibinfo{volume}{122}},
  \bibinfo{pages}{755} (\bibinfo{year}{2004}).

\bibitem[{\citenamefont{Scott and Harding}(1986)}]{Hard86}
\bibinfo{author}{\bibfnamefont{I.~R.} \bibnamefont{Scott}} \bibnamefont{and}
  \bibinfo{author}{\bibfnamefont{C.~R.} \bibnamefont{Harding}},
  \bibinfo{journal}{Dev. Biol.} \textbf{\bibinfo{volume}{115}},
  \bibinfo{pages}{84} (\bibinfo{year}{1986}).

\bibitem[{\citenamefont{Harding}(2004)}]{Hard04}
\bibinfo{author}{\bibfnamefont{C.~R.} \bibnamefont{Harding}},
  \bibinfo{journal}{Dermatol. Ther.} \textbf{\bibinfo{volume}{17}},
  \bibinfo{pages}{6} (\bibinfo{year}{2004}).

\bibitem[{\citenamefont{Rawlings and Harding}(2004)}]{Rawl04}
\bibinfo{author}{\bibfnamefont{A.~V.} \bibnamefont{Rawlings}} \bibnamefont{and}
  \bibinfo{author}{\bibfnamefont{C.~R.} \bibnamefont{Harding}},
  \bibinfo{journal}{Dermatol. Ther.} \textbf{\bibinfo{volume}{17}},
  \bibinfo{pages}{43} (\bibinfo{year}{2004}).

\bibitem[{\citenamefont{Takada et~al.}(2012)\citenamefont{Takada, Naito,
  Sonoda, and Miyauchi}}]{Tak12}
\bibinfo{author}{\bibfnamefont{S.}~\bibnamefont{Takada}},
  \bibinfo{author}{\bibfnamefont{S.}~\bibnamefont{Naito}},
  \bibinfo{author}{\bibfnamefont{J.}~\bibnamefont{Sonoda}}, \bibnamefont{and}
  \bibinfo{author}{\bibfnamefont{Y.}~\bibnamefont{Miyauchi}},
  \bibinfo{journal}{Appl. Spectrosc.} \textbf{\bibinfo{volume}{66}},
  \bibinfo{pages}{26} (\bibinfo{year}{2012}).

\bibitem[{\citenamefont{Jacobson et~al.}(1990)\citenamefont{Jacobson,
  {Y\"uksel}, Geesin, Gordon, Lane, and Gracy}}]{Jac90}
\bibinfo{author}{\bibfnamefont{T.~M.} \bibnamefont{Jacobson}},
  \bibinfo{author}{\bibfnamefont{K.~U.} \bibnamefont{{Y\"uksel}}},
  \bibinfo{author}{\bibfnamefont{J.~C.} \bibnamefont{Geesin}},
  \bibinfo{author}{\bibfnamefont{J.~S.} \bibnamefont{Gordon}},
  \bibinfo{author}{\bibfnamefont{A.~T.} \bibnamefont{Lane}}, \bibnamefont{and}
  \bibinfo{author}{\bibfnamefont{R.~W.} \bibnamefont{Gracy}},
  \bibinfo{journal}{J. Invest. Dermatol.} \textbf{\bibinfo{volume}{95}},
  \bibinfo{pages}{296} (\bibinfo{year}{1990}).

\bibitem[{\citenamefont{Rawlings et~al.}(1994)\citenamefont{Rawlings, Scott,
  Harding, and Bowser}}]{Rawl94}
\bibinfo{author}{\bibfnamefont{A.~V.} \bibnamefont{Rawlings}},
  \bibinfo{author}{\bibfnamefont{I.~R.} \bibnamefont{Scott}},
  \bibinfo{author}{\bibfnamefont{C.~R.} \bibnamefont{Harding}},
  \bibnamefont{and} \bibinfo{author}{\bibfnamefont{P.~A.}
  \bibnamefont{Bowser}}, \bibinfo{journal}{J. Invest. Dermatol.}
  \textbf{\bibinfo{volume}{103}}, \bibinfo{pages}{731} (\bibinfo{year}{1994}).

\bibitem[{\citenamefont{Fleer et~al.}(1993)\citenamefont{Fleer, Cohen~Stuart,
  Scheutjens, Cosgrove, and Vincent}}]{fleer-book}
\bibinfo{author}{\bibfnamefont{G.~J.} \bibnamefont{Fleer}},
  \bibinfo{author}{\bibfnamefont{M.~A.} \bibnamefont{Cohen~Stuart}},
  \bibinfo{author}{\bibfnamefont{J.~M. H.~M.} \bibnamefont{Scheutjens}},
  \bibinfo{author}{\bibfnamefont{T.}~\bibnamefont{Cosgrove}}, \bibnamefont{and}
  \bibinfo{author}{\bibfnamefont{B.}~\bibnamefont{Vincent}},
  \emph{\bibinfo{title}{Polymers at interfaces}}
  (\bibinfo{publisher}{Springer}, \bibinfo{address}{New York},
  \bibinfo{year}{1993}).

\bibitem[{\citenamefont{Israels et~al.}(1993)\citenamefont{Israels, Scheutjens,
  and Fleer}}]{Israels93}
\bibinfo{author}{\bibfnamefont{R.}~\bibnamefont{Israels}},
  \bibinfo{author}{\bibfnamefont{J.~M. H.~M.} \bibnamefont{Scheutjens}},
  \bibnamefont{and} \bibinfo{author}{\bibfnamefont{G.~J.} \bibnamefont{Fleer}},
  \bibinfo{journal}{Macromol.} \textbf{\bibinfo{volume}{26}},
  \bibinfo{pages}{5405} (\bibinfo{year}{1993}).

\bibitem[{\citenamefont{Israels et~al.}(1994)\citenamefont{Israels, Leermakers,
  and Fleer}}]{Israels94}
\bibinfo{author}{\bibfnamefont{R.}~\bibnamefont{Israels}},
  \bibinfo{author}{\bibfnamefont{F.~A.~M.} \bibnamefont{Leermakers}},
  \bibnamefont{and} \bibinfo{author}{\bibfnamefont{G.~J.} \bibnamefont{Fleer}},
  \bibinfo{journal}{Macromol.} \textbf{\bibinfo{volume}{27}},
  \bibinfo{pages}{3087} (\bibinfo{year}{1994}).

\bibitem[{\citenamefont{B\"ohmer et~al.}(1990)\citenamefont{B\"ohmer, Evers,
  and Scheutjens}}]{Bohmer90}
\bibinfo{author}{\bibfnamefont{M.~R.} \bibnamefont{B\"ohmer}},
  \bibinfo{author}{\bibfnamefont{O.~A.} \bibnamefont{Evers}}, \bibnamefont{and}
  \bibinfo{author}{\bibfnamefont{J.~M. H.~M.} \bibnamefont{Scheutjens}},
  \bibinfo{journal}{Macromol.} \textbf{\bibinfo{volume}{23}},
  \bibinfo{pages}{2288} (\bibinfo{year}{1990}).

\bibitem[{\citenamefont{Evers et~al.}(1990{\natexlab{a}})\citenamefont{Evers,
  Scheutjens, and Fleer}}]{Evers1}
\bibinfo{author}{\bibfnamefont{O.~A.} \bibnamefont{Evers}},
  \bibinfo{author}{\bibfnamefont{J.~M. H.~M.} \bibnamefont{Scheutjens}},
  \bibnamefont{and} \bibinfo{author}{\bibfnamefont{G.~J.} \bibnamefont{Fleer}},
  \bibinfo{journal}{Macromol.} \textbf{\bibinfo{volume}{23}},
  \bibinfo{pages}{5221} (\bibinfo{year}{1990}{\natexlab{a}}).

\bibitem[{\citenamefont{Evers et~al.}(1990{\natexlab{b}})\citenamefont{Evers,
  Scheutjens, and Fleer}}]{Evers2}
\bibinfo{author}{\bibfnamefont{O.~A.} \bibnamefont{Evers}},
  \bibinfo{author}{\bibfnamefont{J.~M. H.~M.} \bibnamefont{Scheutjens}},
  \bibnamefont{and} \bibinfo{author}{\bibfnamefont{G.~J.} \bibnamefont{Fleer}},
  \bibinfo{journal}{J. Chem. Soc. Faraday Trans.}
  \textbf{\bibinfo{volume}{86}}, \bibinfo{pages}{1333}
  (\bibinfo{year}{1990}{\natexlab{b}}).

\bibitem[{\citenamefont{Evers et~al.}(1991)\citenamefont{Evers, Scheutjens, and
  Fleer}}]{Evers3}
\bibinfo{author}{\bibfnamefont{O.~A.} \bibnamefont{Evers}},
  \bibinfo{author}{\bibfnamefont{J.~M. H.~M.} \bibnamefont{Scheutjens}},
  \bibnamefont{and} \bibinfo{author}{\bibfnamefont{G.~J.} \bibnamefont{Fleer}},
  \bibinfo{journal}{Macromol.} \textbf{\bibinfo{volume}{24}},
  \bibinfo{pages}{5558} (\bibinfo{year}{1991}).

\bibitem[{\citenamefont{Leermakers et~al.}(1996)\citenamefont{Leermakers,
  Atkinson, Dickinson, and Horne}}]{Leer96}
\bibinfo{author}{\bibfnamefont{F.~A.~M.} \bibnamefont{Leermakers}},
  \bibinfo{author}{\bibfnamefont{P.~J.} \bibnamefont{Atkinson}},
  \bibinfo{author}{\bibfnamefont{E.}~\bibnamefont{Dickinson}},
  \bibnamefont{and} \bibinfo{author}{\bibfnamefont{D.~S.} \bibnamefont{Horne}},
  \bibinfo{journal}{J. Coll. Int. Sci.} \textbf{\bibinfo{volume}{178}},
  \bibinfo{pages}{681} (\bibinfo{year}{1996}).

\bibitem[{\citenamefont{Dickinson
  et~al.}(1997{\natexlab{a}})\citenamefont{Dickinson, Horne, Pinfield, and
  Leermakers}}]{Dick97a}
\bibinfo{author}{\bibfnamefont{E.}~\bibnamefont{Dickinson}},
  \bibinfo{author}{\bibfnamefont{D.~S.} \bibnamefont{Horne}},
  \bibinfo{author}{\bibfnamefont{V.~J.} \bibnamefont{Pinfield}},
  \bibnamefont{and} \bibinfo{author}{\bibfnamefont{F.~A.~M.}
  \bibnamefont{Leermakers}}, \bibinfo{journal}{J. Chem. Soc. Faraday Trans.}
  \textbf{\bibinfo{volume}{93}}, \bibinfo{pages}{425}
  (\bibinfo{year}{1997}{\natexlab{a}}).

\bibitem[{\citenamefont{Dickinson
  et~al.}(1997{\natexlab{b}})\citenamefont{Dickinson, Pinfield, Horne, and
  Leermakers}}]{Dick97b}
\bibinfo{author}{\bibfnamefont{E.}~\bibnamefont{Dickinson}},
  \bibinfo{author}{\bibfnamefont{V.~J.} \bibnamefont{Pinfield}},
  \bibinfo{author}{\bibfnamefont{D.~S.} \bibnamefont{Horne}}, \bibnamefont{and}
  \bibinfo{author}{\bibfnamefont{F.~A.~M.} \bibnamefont{Leermakers}},
  \bibinfo{journal}{J. Chem. Soc. Faraday Trans.}
  \textbf{\bibinfo{volume}{93}}, \bibinfo{pages}{1785}
  (\bibinfo{year}{1997}{\natexlab{b}}).

\bibitem[{\citenamefont{Ettelaie et~al.}(2008)\citenamefont{Ettelaie,
  Akinshina, and Dickinson}}]{Ett08}
\bibinfo{author}{\bibfnamefont{R.}~\bibnamefont{Ettelaie}},
  \bibinfo{author}{\bibfnamefont{A.}~\bibnamefont{Akinshina}},
  \bibnamefont{and}
  \bibinfo{author}{\bibfnamefont{E.}~\bibnamefont{Dickinson}},
  \bibinfo{journal}{Faraday Discuss.} \textbf{\bibinfo{volume}{139}},
  \bibinfo{pages}{161} (\bibinfo{year}{2008}).

\bibitem[{\citenamefont{Akinshina et~al.}(2008)\citenamefont{Akinshina,
  Ettelaie, Dickinson, and Smyth}}]{Akin08}
\bibinfo{author}{\bibfnamefont{A.}~\bibnamefont{Akinshina}},
  \bibinfo{author}{\bibfnamefont{R.}~\bibnamefont{Ettelaie}},
  \bibinfo{author}{\bibfnamefont{E.}~\bibnamefont{Dickinson}},
  \bibnamefont{and} \bibinfo{author}{\bibfnamefont{G.}~\bibnamefont{Smyth}},
  \bibinfo{journal}{Biomacromolecules} \textbf{\bibinfo{volume}{9}},
  \bibinfo{pages}{3188} (\bibinfo{year}{2008}).

\bibitem[{\citenamefont{Zhulina and Leermakers}(2007{\natexlab{a}})}]{Zhu07a}
\bibinfo{author}{\bibfnamefont{E.~B.} \bibnamefont{Zhulina}} \bibnamefont{and}
  \bibinfo{author}{\bibfnamefont{F.~A.~M.} \bibnamefont{Leermakers}},
  \bibinfo{journal}{Biophys. J.} \textbf{\bibinfo{volume}{93}},
  \bibinfo{pages}{1421} (\bibinfo{year}{2007}{\natexlab{a}}).

\bibitem[{\citenamefont{Zhulina and Leermakers}(2007{\natexlab{b}})}]{Zhu07b}
\bibinfo{author}{\bibfnamefont{E.~B.} \bibnamefont{Zhulina}} \bibnamefont{and}
  \bibinfo{author}{\bibfnamefont{F.~A.~M.} \bibnamefont{Leermakers}},
  \bibinfo{journal}{Biophys. J.} \textbf{\bibinfo{volume}{93}},
  \bibinfo{pages}{1452} (\bibinfo{year}{2007}{\natexlab{b}}).

\bibitem[{\citenamefont{Zhulina and Leermakers}(2009)}]{Zhu09}
\bibinfo{author}{\bibfnamefont{E.~B.} \bibnamefont{Zhulina}} \bibnamefont{and}
  \bibinfo{author}{\bibfnamefont{F.~A.~M.} \bibnamefont{Leermakers}},
  \bibinfo{journal}{Soft Matter} \textbf{\bibinfo{volume}{5}},
  \bibinfo{pages}{2836} (\bibinfo{year}{2009}).

\bibitem[{\citenamefont{Zhulina and Leermakers}(2010)}]{Zhu10}
\bibinfo{author}{\bibfnamefont{E.~B.} \bibnamefont{Zhulina}} \bibnamefont{and}
  \bibinfo{author}{\bibfnamefont{F.~A.~M.} \bibnamefont{Leermakers}},
  \bibinfo{journal}{Biophys. J.} \textbf{\bibinfo{volume}{98}},
  \bibinfo{pages}{462} (\bibinfo{year}{2010}).

\bibitem[{\citenamefont{Leermakers and Zhulina}(2010)}]{Leer10a}
\bibinfo{author}{\bibfnamefont{F.~A.~M.} \bibnamefont{Leermakers}}
  \bibnamefont{and} \bibinfo{author}{\bibfnamefont{E.~B.}
  \bibnamefont{Zhulina}}, \bibinfo{journal}{Eur. Biophys. J.}
  \textbf{\bibinfo{volume}{39}}, \bibinfo{pages}{1323} (\bibinfo{year}{2010}).

\bibitem[{\citenamefont{Leermakers et~al.}(2010)\citenamefont{Leermakers, Jho,
  and Zhulina}}]{Leer10b}
\bibinfo{author}{\bibfnamefont{F.~A.~M.} \bibnamefont{Leermakers}},
  \bibinfo{author}{\bibfnamefont{Y.-S.} \bibnamefont{Jho}}, \bibnamefont{and}
  \bibinfo{author}{\bibfnamefont{E.~B.} \bibnamefont{Zhulina}},
  \bibinfo{journal}{Soft Matter} \textbf{\bibinfo{volume}{6}},
  \bibinfo{pages}{5533} (\bibinfo{year}{2010}).

\bibitem[{if-()}]{if-note}
\bibinfo{note}{See http://www.interfil.org/.}

\bibitem[{\citenamefont{Ohnishi et~al.}(2006)\citenamefont{Ohnishi, Kamikubo,
  Onitsuka, Kataoka, and Shortle}}]{Ohn06}
\bibinfo{author}{\bibfnamefont{S.}~\bibnamefont{Ohnishi}},
  \bibinfo{author}{\bibfnamefont{H.}~\bibnamefont{Kamikubo}},
  \bibinfo{author}{\bibfnamefont{M.}~\bibnamefont{Onitsuka}},
  \bibinfo{author}{\bibfnamefont{M.}~\bibnamefont{Kataoka}}, \bibnamefont{and}
  \bibinfo{author}{\bibfnamefont{D.}~\bibnamefont{Shortle}},
  \bibinfo{journal}{J. Am. Chem. Soc.} \textbf{\bibinfo{volume}{128}},
  \bibinfo{pages}{16338} (\bibinfo{year}{2006}).

\bibitem[{\citenamefont{Bykov and Asher}(2010)}]{Byk10}
\bibinfo{author}{\bibfnamefont{S.}~\bibnamefont{Bykov}} \bibnamefont{and}
  \bibinfo{author}{\bibfnamefont{S.}~\bibnamefont{Asher}}, \bibinfo{journal}{J.
  Phys. Chem. B} \textbf{\bibinfo{volume}{114}}, \bibinfo{pages}{6636}
  (\bibinfo{year}{2010}).

\bibitem[{\citenamefont{Nozaki and Tanford}(1971)}]{Noz71}
\bibinfo{author}{\bibfnamefont{Y.}~\bibnamefont{Nozaki}} \bibnamefont{and}
  \bibinfo{author}{\bibfnamefont{C.}~\bibnamefont{Tanford}},
  \bibinfo{journal}{J. Biol. Chem.} \textbf{\bibinfo{volume}{246}},
  \bibinfo{pages}{2211} (\bibinfo{year}{1971}).

\bibitem[{\citenamefont{Lu et~al.}(2006)\citenamefont{Lu, Wang, Yang, and
  Ching}}]{Lu06}
\bibinfo{author}{\bibfnamefont{J.}~\bibnamefont{Lu}},
  \bibinfo{author}{\bibfnamefont{X.-J.} \bibnamefont{Wang}},
  \bibinfo{author}{\bibfnamefont{X.}~\bibnamefont{Yang}}, \bibnamefont{and}
  \bibinfo{author}{\bibfnamefont{C.-B.} \bibnamefont{Ching}},
  \bibinfo{journal}{J. Chem. Eng. Data} \textbf{\bibinfo{volume}{51}},
  \bibinfo{pages}{1593} (\bibinfo{year}{2006}).

\bibitem[{\citenamefont{Hanson et~al.}(2002)\citenamefont{Hanson, Behne, Barry,
  Mauro, Gratton, and Clegg}}]{Hanson02}
\bibinfo{author}{\bibfnamefont{K.~M.} \bibnamefont{Hanson}},
  \bibinfo{author}{\bibfnamefont{M.~J.} \bibnamefont{Behne}},
  \bibinfo{author}{\bibfnamefont{N.~P.} \bibnamefont{Barry}},
  \bibinfo{author}{\bibfnamefont{T.~M.} \bibnamefont{Mauro}},
  \bibinfo{author}{\bibfnamefont{E.}~\bibnamefont{Gratton}}, \bibnamefont{and}
  \bibinfo{author}{\bibfnamefont{R.~M.} \bibnamefont{Clegg}},
  \bibinfo{journal}{Biophys. J.} \textbf{\bibinfo{volume}{83}},
  \bibinfo{pages}{1682} (\bibinfo{year}{2002}).

\bibitem[{\citenamefont{Parra and Paye}(2003)}]{Parra03}
\bibinfo{author}{\bibfnamefont{J.~L.} \bibnamefont{Parra}} \bibnamefont{and}
  \bibinfo{author}{\bibfnamefont{M.}~\bibnamefont{Paye}},
  \bibinfo{journal}{Skin Pharmacol. Appl.} \textbf{\bibinfo{volume}{16}},
  \bibinfo{pages}{188} (\bibinfo{year}{2003}).

\bibitem[{\citenamefont{\AA{}berg et~al.}(2008)\citenamefont{\AA{}berg,
  Wennerstr\"om, and Sparr}}]{Aberg08}
\bibinfo{author}{\bibfnamefont{C.}~\bibnamefont{\AA{}berg}},
  \bibinfo{author}{\bibfnamefont{H.}~\bibnamefont{Wennerstr\"om}},
  \bibnamefont{and} \bibinfo{author}{\bibfnamefont{E.}~\bibnamefont{Sparr}},
  \bibinfo{journal}{Langmuir} \textbf{\bibinfo{volume}{24}},
  \bibinfo{pages}{8061} (\bibinfo{year}{2008}).

\bibitem[{\citenamefont{Carta}(1999)}]{Car99}
\bibinfo{author}{\bibfnamefont{R.}~\bibnamefont{Carta}}, \bibinfo{journal}{J.
  Chem. Eng. Data} \textbf{\bibinfo{volume}{44}}, \bibinfo{pages}{563}
  (\bibinfo{year}{1999}).

\bibitem[{trp()}]{trp-note}
\bibinfo{note}{See http://www.imb-jena.de/IMAGE\_AA.html.}

\bibitem[{\citenamefont{Fuchs and Weber}(1994)}]{Fuchs94}
\bibinfo{author}{\bibfnamefont{E.}~\bibnamefont{Fuchs}} \bibnamefont{and}
  \bibinfo{author}{\bibfnamefont{K.}~\bibnamefont{Weber}},
  \bibinfo{journal}{Annu. Rev. Biochem.} \textbf{\bibinfo{volume}{63}},
  \bibinfo{pages}{345} (\bibinfo{year}{1994}).

\bibitem[{\citenamefont{Steinert et~al.}(1985)\citenamefont{Steinert, Parr,
  Idler, Johnson, Steven, and Roop}}]{Stein85a}
\bibinfo{author}{\bibfnamefont{P.~M.} \bibnamefont{Steinert}},
  \bibinfo{author}{\bibfnamefont{y.~D.~A.} \bibnamefont{Parr}},
  \bibinfo{author}{\bibfnamefont{W.~W.} \bibnamefont{Idler}},
  \bibinfo{author}{\bibfnamefont{L.~D.} \bibnamefont{Johnson}},
  \bibinfo{author}{\bibfnamefont{A.~C.} \bibnamefont{Steven}},
  \bibnamefont{and} \bibinfo{author}{\bibfnamefont{D.~R.} \bibnamefont{Roop}},
  \bibinfo{journal}{J. Biol. Chem.} \textbf{\bibinfo{volume}{260}},
  \bibinfo{pages}{7142} (\bibinfo{year}{1985}).

\bibitem[{\citenamefont{Steinert et~al.}(1991)\citenamefont{Steinert, Mack,
  Korge, Gan, Haynes, and Steven}}]{Stein91}
\bibinfo{author}{\bibfnamefont{P.~M.} \bibnamefont{Steinert}},
  \bibinfo{author}{\bibfnamefont{J.~W.} \bibnamefont{Mack}},
  \bibinfo{author}{\bibfnamefont{B.~P.} \bibnamefont{Korge}},
  \bibinfo{author}{\bibfnamefont{S.~Q.} \bibnamefont{Gan}},
  \bibinfo{author}{\bibfnamefont{S.~R.} \bibnamefont{Haynes}},
  \bibnamefont{and} \bibinfo{author}{\bibfnamefont{A.~C.}
  \bibnamefont{Steven}}, \bibinfo{journal}{Int. J. Biol. Macromol.}
  \textbf{\bibinfo{volume}{13}}, \bibinfo{pages}{130} (\bibinfo{year}{1991}).

\bibitem[{\citenamefont{Israelachvili}(2011)}]{israel-book}
\bibinfo{author}{\bibfnamefont{J.~N.} \bibnamefont{Israelachvili}},
  \emph{\bibinfo{title}{Intermolecular and surface forces}}
  (\bibinfo{publisher}{Academic Press}, \bibinfo{address}{Burlington},
  \bibinfo{year}{2011}).

\bibitem[{\citenamefont{Scheutjens and Fleer}(1985)}]{SF85}
\bibinfo{author}{\bibfnamefont{J.~M. H.~M.} \bibnamefont{Scheutjens}}
  \bibnamefont{and} \bibinfo{author}{\bibfnamefont{G.~J.} \bibnamefont{Fleer}},
  \bibinfo{journal}{Macromol.} \textbf{\bibinfo{volume}{18}},
  \bibinfo{pages}{1882} (\bibinfo{year}{1985}).

\bibitem[{\citenamefont{Miklavic et~al.}(1990)\citenamefont{Miklavic, Woodward,
  {J\"onsson}, and {\AA{}kesson}}}]{Mik90}
\bibinfo{author}{\bibfnamefont{S.~J.} \bibnamefont{Miklavic}},
  \bibinfo{author}{\bibfnamefont{C.~E.} \bibnamefont{Woodward}},
  \bibinfo{author}{\bibfnamefont{B.}~\bibnamefont{{J\"onsson}}},
  \bibnamefont{and}
  \bibinfo{author}{\bibfnamefont{T.}~\bibnamefont{{\AA{}kesson}}},
  \bibinfo{journal}{Macromolecules} \textbf{\bibinfo{volume}{23}},
  \bibinfo{pages}{4149} (\bibinfo{year}{1990}).

\bibitem[{\citenamefont{Warner et~al.}(1988)\citenamefont{Warner, Myers, and
  Taylor}}]{Warner88}
\bibinfo{author}{\bibfnamefont{R.~R.} \bibnamefont{Warner}},
  \bibinfo{author}{\bibfnamefont{M.~C.} \bibnamefont{Myers}}, \bibnamefont{and}
  \bibinfo{author}{\bibfnamefont{D.~A.} \bibnamefont{Taylor}},
  \bibinfo{journal}{J. Invest. Dermatol.} \textbf{\bibinfo{volume}{90}},
  \bibinfo{pages}{218} (\bibinfo{year}{1988}).

\bibitem[{\citenamefont{Caspers et~al.}(2001)\citenamefont{Caspers, Lucassen,
  Carter, Bruining, and Puppels}}]{Caspers01}
\bibinfo{author}{\bibfnamefont{P.~J.} \bibnamefont{Caspers}},
  \bibinfo{author}{\bibfnamefont{G.~W.} \bibnamefont{Lucassen}},
  \bibinfo{author}{\bibfnamefont{E.~A.} \bibnamefont{Carter}},
  \bibinfo{author}{\bibfnamefont{H.~A.} \bibnamefont{Bruining}},
  \bibnamefont{and} \bibinfo{author}{\bibfnamefont{G.~J.}
  \bibnamefont{Puppels}}, \bibinfo{journal}{J. Invest. Dermatol.}
  \textbf{\bibinfo{volume}{116}}, \bibinfo{pages}{434} (\bibinfo{year}{2001}).

\bibitem[{\citenamefont{Claesson and Ninham}(1992)}]{Clae92}
\bibinfo{author}{\bibfnamefont{P.}~\bibnamefont{Claesson}} \bibnamefont{and}
  \bibinfo{author}{\bibfnamefont{B.}~\bibnamefont{Ninham}},
  \bibinfo{journal}{Langmuir} \textbf{\bibinfo{volume}{8}},
  \bibinfo{pages}{1406} (\bibinfo{year}{1992}).

\bibitem[{\citenamefont{Dahlgren et~al.}(1993)\citenamefont{Dahlgren, Waltermo,
  Blomberg, Claesson, {Sj\"ostr\"om}, {\AA{}kesson}, and
  {J\"onsson}}}]{Dahlgren93}
\bibinfo{author}{\bibfnamefont{M.~A.~G.} \bibnamefont{Dahlgren}},
  \bibinfo{author}{\bibfnamefont{{\AA}.}~\bibnamefont{Waltermo}},
  \bibinfo{author}{\bibfnamefont{E.}~\bibnamefont{Blomberg}},
  \bibinfo{author}{\bibfnamefont{P.~M.} \bibnamefont{Claesson}},
  \bibinfo{author}{\bibfnamefont{L.}~\bibnamefont{{Sj\"ostr\"om}}},
  \bibinfo{author}{\bibfnamefont{T.}~\bibnamefont{{\AA{}kesson}}},
  \bibnamefont{and}
  \bibinfo{author}{\bibfnamefont{B.}~\bibnamefont{{J\"onsson}}},
  \bibinfo{journal}{J. Phys. Chem.} \textbf{\bibinfo{volume}{97}},
  \bibinfo{pages}{11769} (\bibinfo{year}{1993}).

\bibitem[{\citenamefont{Borukhov et~al.}(1999)\citenamefont{Borukhov, Andelman,
  and Orland}}]{Borukhov99}
\bibinfo{author}{\bibfnamefont{I.}~\bibnamefont{Borukhov}},
  \bibinfo{author}{\bibfnamefont{D.}~\bibnamefont{Andelman}}, \bibnamefont{and}
  \bibinfo{author}{\bibfnamefont{H.}~\bibnamefont{Orland}},
  \bibinfo{journal}{J. Phys. Chem. B} \textbf{\bibinfo{volume}{103}},
  \bibinfo{pages}{5042} (\bibinfo{year}{1999}).

\end{thebibliography}

\end{document}